# Pseudorapidity density, transverse momentum spectra, and elliptic flow studies in Xe-Xe collision systems at $\sqrt{s_{NN}} = 5.44$ TeV using the HYDJET++ model

Saraswati Pandey,[1,*] S. K. Tiwari,[2,†] and B. K. Singh[1,‡]
[1]*Department of Physics, Institute of Science, Banaras Hindu University, Varanasi 221005, India*
[2]*Department of Applied Science and Humanities, MIT, Muzaffarpur 842003, Bihar, India*

(Received 5 August 2020; accepted 26 October 2020; published 14 January 2021)

In this paper, we present a systematic study of Xe-Xe collisions at 5.44-TeV center-of-mass energy. We employ the Monte Carlo (hydrodynamics plus jets) HYDJET++ model to calculate the pseudorapidity distribution, transverse momentum ($p_T$) spectra and the elliptic flow ($v_2$) of charged hadrons with different parameters in two geometrical configurations: body-body and tip-tip type of Xe-Xe collisions. The kinematic ranges $0 < p_T < 50$ GeV/$c$ and $|\eta| < 0.8$ are considered for our paper. Results are obtained for seven classes of centrality. For comparison, we have shown results from the AMPT model with the string-melting version. The results obtained for Xe-Xe collision systems for minimum bias at midrapidity match well with the experimental data of the ALICE/CMS Collaborations. We observe that the pseudorapidity density depends on the size and geometry of the colliding system. The centrality dependence of average transverse momentum ($\langle p_T \rangle$) and average elliptic flow ($\langle v_2 \rangle$) is also observed. The charged hadron properties also show clear dependence on the geometrical configuration of the collisions. Our model results have been compared to results obtained in the AMPT model. The HYDJET++ model justifies data more closely than the AMPT model, the latter mostly overpredicts the experimental data.

## I. INTRODUCTION

One of the main objectives of various heavy-ion collision programs running at various collider machines, such as the Relativistic Heavy-Ion collider (RHIC) at Brookhaven National Laboratory and the Large Hadron Collider (LHC) at CERN is to study the matter at extreme conditions of temperature and density. Quantum chromodynamics, an established theory of strong interactions predicts a phase transition from hadron gas to quark-gluon plasma which can best be studied in such collision programs [1]. This plasma of strongly interacting quarks and gluons is formed in the hot and dense nuclear matter (usually called the fireball) created in such heavy-ion collisions experiments. One of the main motivations is to understand the properties of the quark-gluon plasma (QGP) by analyzing its typical signatures, such as elliptic flow [2], jet quenching [3], high-$p_T$ suppression [4,5], etc. These signatures depend upon the basic macroscopic properties, such as the shape and size of the QGP formed in heavy-ion collisions. So, we can study the evolution of these observables in the ambit of varying collision geometry. One way to accomplish this is by colliding different species significantly varying by their mass numbers. Another way may include, the study of a particular colliding system by dividing them into different classes of centrality. We choose here the latter one for studying and analyzing particle production, transverse momentum distribution, and elliptic flow of the charged particles in our paper. Since the overall particle production in the collisions is related to the initial density, the multiplicity of charged particles becomes an important observable to characterize the properties of dense matter created in the collisions. Also, the transverse momentum spectra of the charged particles carry essential information of the QGP formed in the collision systems [6]. Anisotropic flow, an important signature of QGP, comes from the initial asymmetries in the geometry of the system produced in any noncentral collision. It plays an important role in the understanding of the collective motion and the bulk properties of the QGP. Elliptic flow gives a better understanding of the matter formed in a heavy-ion collision system as it is sensitive to the properties of the system at an early stage of its evolution. By having a clear picture of elliptic flow, we can have in-depth information about the equation of state, initial geometrical anisotropies, and the transport properties of the created QGP medium [7]. Furthermore, heavy-ion collisions with deformed nucleus will be a reliable tool to handle background measurements for the chiral magnetic effect (CME) [8,9] which is caused by the magnetic field produced due to the spectator nucleons. Here, the elliptic flow acts as a background effect for the CME observable. In central collisions, both CME and $v_2$ are very low whereas in midcentral collisions both are very high. So, the most suitable collisions to observe CME are the ones where the magnetic field is high or finite and $v_2$ is minimum or negligible. Such a scenario can be obtained in most central collisions in different geometrical configurations thereby [10], having a finite difference between the two. Hence, the background elliptic flow can be separated from the CME signal.

*saraswati.pandey13@bhu.ac.in
†sktiwari4bhu@gmail.com
‡bksingh@bhu.ac.in

Recently, in 2017, the LHC carried out an 8-h run of Xe-Xe collisions at $\sqrt{s_{NN}} = 5.44$-TeV center-of-mass energy. The mass number of the Xe nucleus lies in mid-between $p$ and $Pb^{208}$. Thus, collisions of $Xe^{129}$ nuclei will bridge the gap between the larger Pb-ion systems and the smaller systems, such as $p + p$ and $p +$ Pb. Unlike the Pb nucleus which is spherical in shape, Xe is somewhat prolate [11]. This deformation in xenon allows us to probe a different initial condition. Also, due to nonspherical shape, we can have multiple configurations in terms of the way nuclei collide with each other. But here we will mainly focus on two geometrical configurations: body-body and tip-tip configurations, depending upon the angle, the colliding nuclei make with the reaction plane [12]. Therefore, the overlapping region formed in the Xe-Xe collision will surely not be circular and, thus, the observables will show an effective change when measured compared to the case of nondeformed collisions. In deformed nucleus collisions, the charged particle multiplicity density in the transverse phase space is expected to be higher compared to the spherical or nondeformed nucleus collisions [13,14]. Hydrodynamical models predict an increase in elliptic flow ($v_2$) by 10% for a deformed Xe nucleus compared to the spherical Xe nucleus in central collisions [15].

To bring out our desired paper, we have used the HYDJET++ model. Most of the existing models either involve high-$p_T$ particle production from jet fragmentation or involve production of low-$p_T$ hadrons using thermal statistical processes. Recently, a multiphase transport model (AMPT) was used to study Xe-Xe collisions at 5.44 TeV [16]. However, it overpredicted the charged-particle multiplicity for most central collisions measured in experiments. Similar observations were performed in another work [17] where there the AMPT model in the string-melting version with an improved quark coalescence method was used to predict the charged-particle pseudorapidity multiplicity density at the LHC energies. Here, $dN_{ch}/d\eta$ systematically overestimated by 20% at different centralities. In Ref. [18], the modified version of the wounded quark model is used to study the multiparticle production in Xe-Xe collisions. The HYDJET++ model [19,20] consistently includes the production of hard as well as soft-$p_T$ hadrons to handle the Xe-Xe collision system. It uses the PYTHIA-type initial condition for the hard part and Glauber-type initial condition for the soft part to simulate the Xe-Xe collisions at $\sqrt{s_{NN}} = 5.44$ TeV. In a recent work, the HYDJET++ model was used to study U-U collisions at the 193-GeV center-of-mass energy in body-body and tip-tip geometrical configurations [21].

In this paper, we have studied the centrality dependence of pseudorapidity density, transverse momentum spectra, average transverse momentum, and elliptic flow distribution of charged hadrons in Xe-Xe collisions at the 5.44-TeV center-of-mass energy. We have analyzed our results in body-body and tip-tip geometrical configurations using the HYDJET++ model. Also, we have shown the effect of deformation on the various QGP observables. The paper is organized as follows: In Sec. II, we have briefly discussed formulation of the HYDJET++ model and the incorporation of deformation in the body of the model. In Sec. III, we show the results and discussions part for the pseudorapidity distribution, transverse momentum distribution, and elliptic flow distributions. Lastly, we have summarized our findings in Sec. IV drawn from this paper.

## II. MODEL FORMALISM

HYDJET++ is a Monte Carlo event generator for simulation of relativistic heavy-ion collisions. It functions by superimposing the soft hydro type state and the hard state resulting from the multiparton fragmentation at the same time treating both states independently. It involves thorough treatment of soft hadroproduction (the collective flow phenomenon and the resonance decays) as well as hard parton production and appraises the known medium effects (jet quenching and nuclear shadowing). The imbibed physics of this model and its simulation procedure can be seen in the corresponding article [19,22]. Some of the essentials of this model are as follows:

The model for the hard multiparton part of the HYDJET++ is similar to that of the HYDJET event generator whose details can be found in Refs. [23–25]. The hard state of a HYDJET++ event is treated using the PYTHIA quenched(PYQUEN) model [23] which repairs a jet event that PYTHIA produces by generating binary nucleonic collision vertices according to the Glauber model at a certain impact parameter. PYTHIA is an event generator used to simulate hard nucleon-nucleon (NN) collision with the consideration of only those events whose generated total transverse momentum is higher than $\mathbf{p}_T^{\min}$. $p_T^{\min}$ here is an important parameter that separates the soft part of the event from the hard part. Events for which generated total transverse momentum is less than $p_T^{\min}$ are taken over by the soft part of the model. This event-by-event simulation is further carried out by the rescattering-by-rescattering simulation of the parton path in the dense medium including the radiative and collision energy losses [26–30] per rescattering. Then the final hadronization is performed using the Lund string model [31] for hard partons and in medium emitted gluons. Nuclear shadowing of the parton distribution function is also incorporated using an impact parameter-dependent parametrization obtained in the framework of the Glauber-Gribov theory [32,33].

A notable fact here is that in the HYDJET++ model, nuclear shadowing corrections are implemented by correcting the contribution of the initial coherent multiple scattering in an effective way instead of modifying the parton showering of single NN collisions in PYTHIA. This nuclear effect restricts the number of partons in the incoming hadronic wave function of both nuclei thereby reducing the total jet production cross section. The decrement caused affects the kinematic variables of incoming hard partons and, hence, both initial and final parton momentum spectra are modified as a result of nuclear shadowing.

The soft part of the HYDJET++ event is the thermal hadronic state generated on the chemical and thermal freeze-out hypersurfaces obtained from a parametrization of relativistic hydrodynamics with preset freeze-out conditions. Furthermore, details of the physics frameworks can be found in the corresponding papers [34,35]. In the statistical fast Monte Carlo model for the soft part, particles are generated on a chemical or thermal freeze-out hypersurface obtained by the parametrization or the numerical solution of relativistic hydrodynamics with given initial conditions and the equation of state [36–39]. The main assumption of this model is that the hadronic matter created in a nuclear collision reaches a local equilibrium after a short period of time ($< 1$ fm$/c$) and then expands hydrodynamically.

The particle densities at the chemical freeze-out stage may be too high to consider the particles as free streaming [40], therefore, the presumption of common chemical thermal freeze-outs is hard for justification. Hence, we consider a more knotty scenario in the HYDJET++ with different chemical and thermal freeze-outs ($T_{\rm ch} \geqslant T_{\rm th}$). In between the two freeze-outs the system expands hydrodynamically with frozen chemical composition, cools down, and the hadrons stream freely as soon as the thermal freeze-out temperature is reached.

Originally, the HYDJET++ model does not include the nuclear density profiles for nuclei having deformation. So, the nuclear density function needs to be altered. The Woods-Saxon nuclear density profile function for nondeformed nuclei, such as Pb and Au appears as follows:

$$\rho(r, z, \theta) = \frac{\rho_0}{1 + \exp\frac{(r-R)}{a}}, \quad (1)$$

where, $\rho_0 = \rho_0^{\rm const} = \frac{M}{V} = \frac{3A}{4R^3}$, $R = R_0 A^{1/3}$, where $R_0 = 1.15$ fm.

Hence, for the Xe nucleus, which is moderately deformed, we have modified the nuclear density profile function to incorporate the deformation. The deformed Woods-Saxon nuclear density profile function [41] in spherical polar coordinates is expressed as

$$\rho(r, z, \theta) = \frac{\rho_0}{1 + \exp\frac{[r - R(1+\beta_2 Y_{20} + \beta_4 Y_{40})]}{a}}, \quad (2)$$

where $\rho_0 = \rho_0^{\rm const} + \text{correction}$, $\rho_0^{\rm const} = \frac{M}{V} = \frac{3A}{4R_A^3} = \frac{3A}{4R_l^3}$, $R_A = R(1 + \beta_2 Y_{20} + \beta_4 Y_{40})$, $R_l = R_0(1 + \beta_2 Y_{20} + \beta_4 Y_{40})$, $R = R_0 A^{1/3}$, where $R_0 = 1.15$ fm. The correction term is calculated as $= \rho_0^{\rm const}(\Pi f/R_A)^2$, where $f = 0.54$ fm, $\beta_2 = 0.162$, and $\beta_4 = -0.003$ are the deformation parameters, $a =$ diffuseness parameter $= 0.59$ fm, $Y_{20} = \sqrt{\frac{5}{16\Pi}}(3\cos 2\theta - 1)$, and $Y_{40} = \frac{3}{16\sqrt{\Pi}}(35\cos 4\theta - 30\cos 2\theta + 3)$ are the spherical harmonics. The values of different parameters have been taken from Ref. [11].

However, incorporating the Woods-Saxon nuclear density profile function in HYDJET++ is onerous as the latter works in cylindrical polar coordinates, unlike the AMPT model that works in the spherical polar coordinate system. The conversion of the deformed nuclear density profile function from spherical polar coordinates $(r, \theta, \phi)$ to cylindrical polar coordinates $(\rho, z, \psi)$ is performed by using the relation $\theta = \tan^{-1}(z/r)$ and $\theta = \tan^{-1}(r/z)$ for body-body and tip-tip configurations, respectively. Here $r$ is $\rho$ of the cylindrical polar coordinates, not to cause any confusion with that of the spherical polar coordinates. The Woods-Saxon nuclear density profile in tip- and body-type geometrical configurations can be visualized in Fig. 1 showing nuclear density profiles for various geometrical configurations.

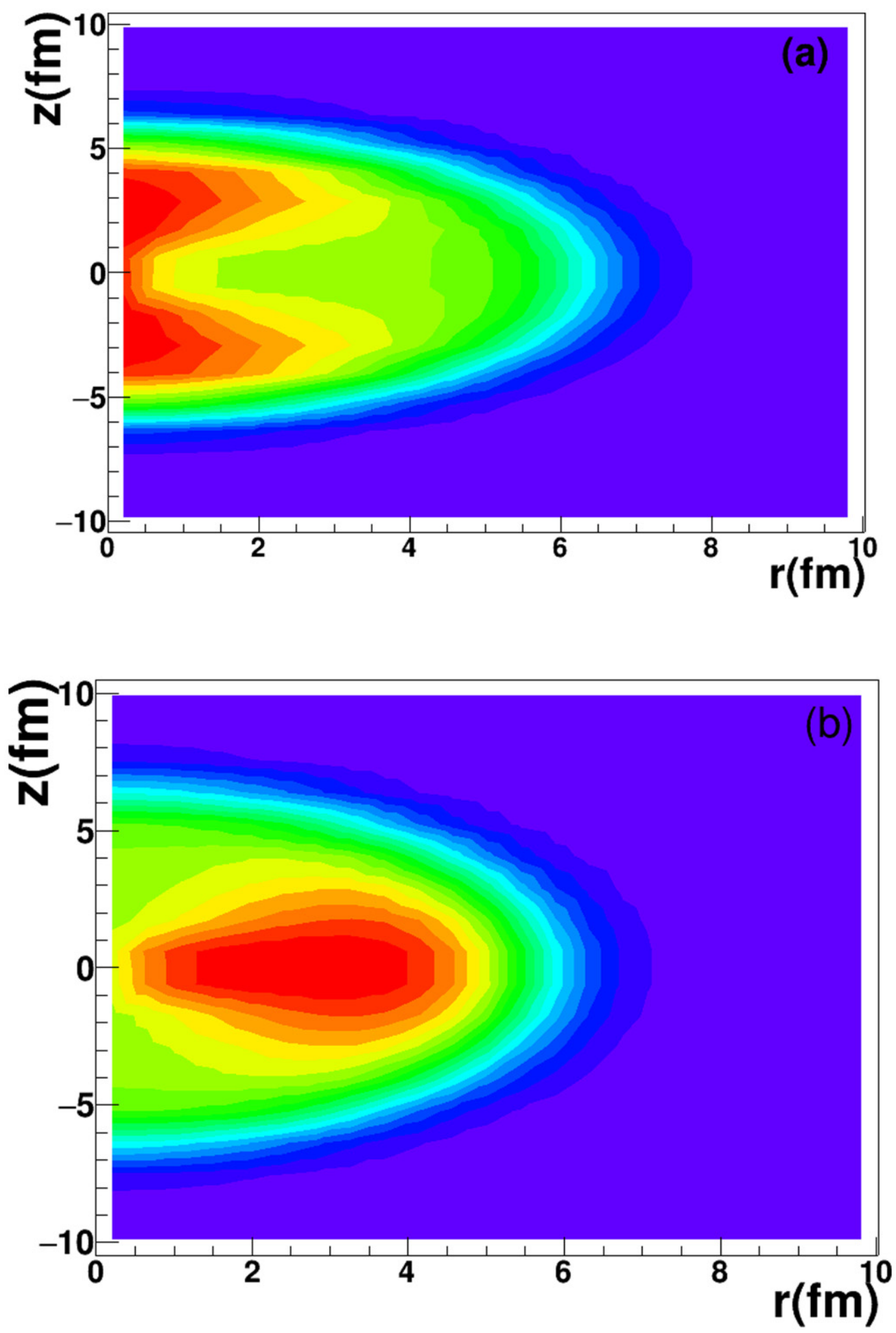

FIG. 1. Woods-Saxon nuclear density profile for xenon in (a) body- and (b) tip-type geometrical configurations under the HYDJET++ framework.

In Fig. 2, we show the variation of the number of the participants and the number of binary collisions in both tip-tip and body-body configurations with respect to the centrality of the events thereby certifying our model. Furthermore, the model parameters which control the execution of our Monte Carlo generator are found by simulating (0–5)% most central Xe-Xe collisions and matching it with the experimental data. The values of different input parameters are provided in Table I,

$$T_0(b, \tau_0) = T_0(b = 0, \tau_0)\left(\frac{N_{\rm part}(b)}{N_{\rm part}(0)}\right)^{1/3}. \quad (3)$$

The kinetic freeze-out temperature for most central collisions is found to be $109 \pm 12$ MeV [42] whereas the chemical freeze-out temperature $T_{\rm ch}$ for most central (0–5)% collisions is around $154 \pm 8$ MeV [42,43]. The temperature for other centralities is calculated using relation (3) to convert the fixed freeze-out hypersurface into a centrality-dependent hypersurface. In this way, we modify the soft particle production in HYDJET++. Furthermore, the hadron multiplicities are calculated using the effective thermal volume approximation and Poisson multiplicity distribution around its mean value which expectedly is proportional to the number of participating nucleons $N_{\text{part}}$ at a given impact parameter. This can be visualized in Fig. 3. The first part in the plot shows the dependence of the chemical freeze-out temperature on the centrality of the events. The second part shows the change in the effective thermal volume with respect to the centrality of the event. We have plotted these for both tip-tip and body-body configurations.

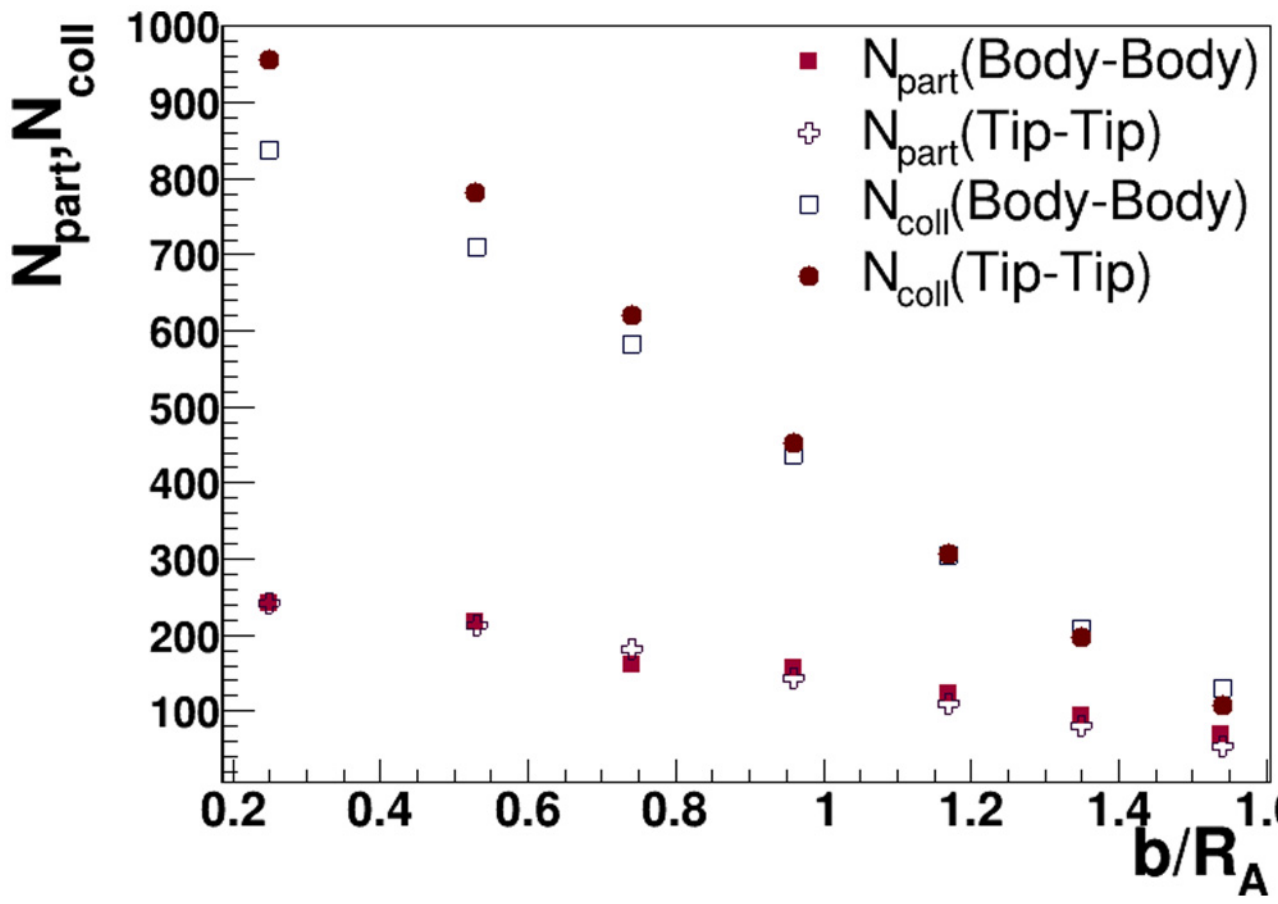

FIG. 2. Variation of the number of participants $N_{\text{part}}$ and the number of binary collisions $N_{\text{coll}}$ with centrality in Xe-Xe collisions.

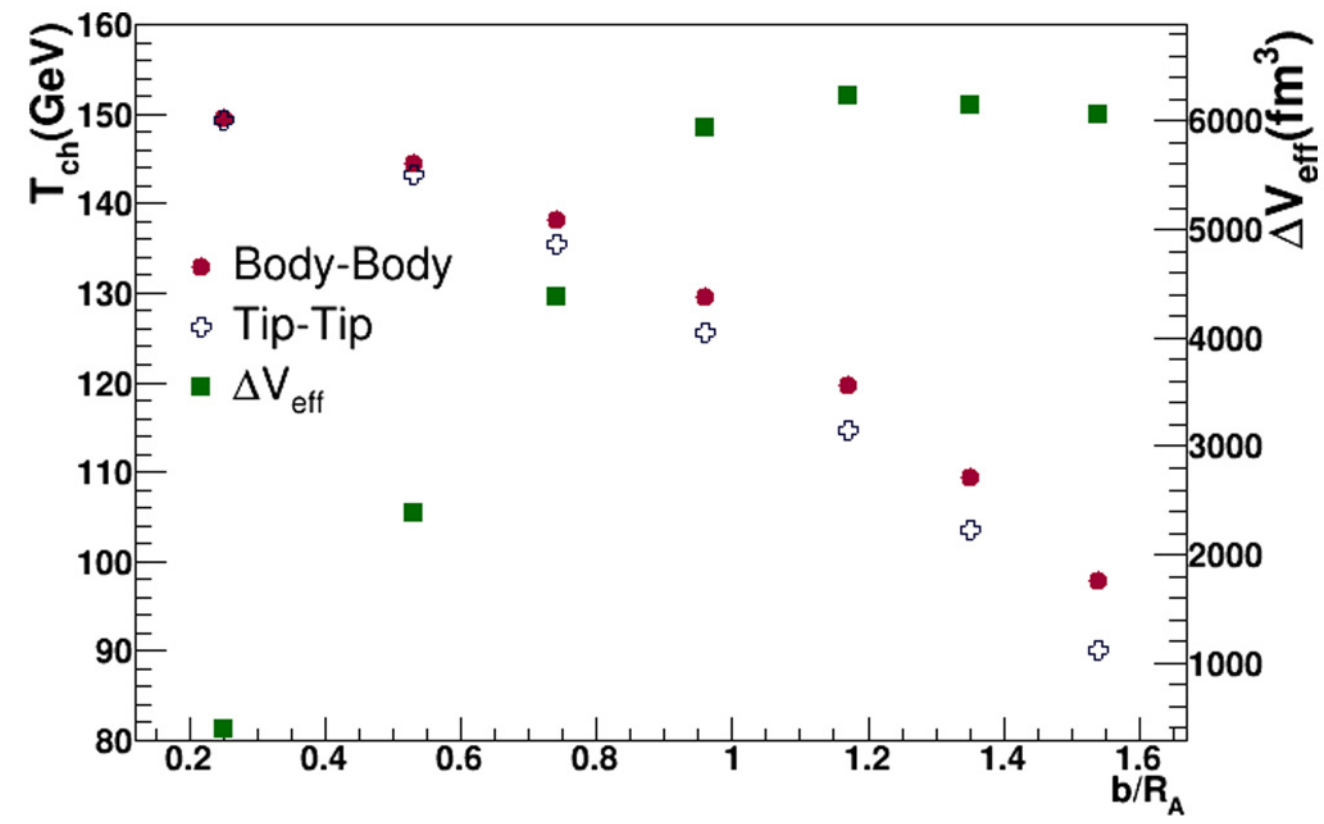

FIG. 3. Variation of $T_{\text{ch}}(b)$ and $\Delta V_{\text{eff}}$ with centrality in Xe-Xe collisions at 5.44 TeV.

## III. RESULTS AND DISCUSSIONS

We have generated $10^5$ events using the modified HYDJET++ model in different centrality classes for both tip-tip and body-body configurations at 5.44-TeV center-of-mass energy and compared the results with the experimental data of the ALICE/CMS Collaborations for our analysis. Only those events have been considered for the results which fall in the kinematic range $|\eta| < 0.8$ and $0 < p_T < 50$ GeV/$c$. Also, we have compared our results with those of the AMPT model in the string-melting version with $|\eta| < 0.5$ pseudorapidity cut [16].

### A. Pseudorapidity distributions

In Fig. 4 we present the comparison of the original HYDJET++ results for Xe-Xe collisions at 5.44 TeV with the modified HYDJET++ results. For completeness, we have also shown the original HYDJET++ results for Pb-Pb collisions at $\sqrt{s_{NN}} = 5.02$ TeV. The results have been compared with the experimental data of the CMS/ALICE Collaborations [44,45]. The original HYDJET++ results for most central Pb-Pb collisions at 5.02 TeV match well with the experimen-tal data of the ALICE Collaboration [45] in the midrapidity region. However, the modified HYDJET++ results show a better match with the experimental data of the CMS Collaboration [44] for midrapidity region than original HYDJET++ ones in Xe- Xe collisions at $\sqrt{s_{NN}} = 5.44$ TeV. This can be very clearly seen in part (b) of the figure. The HYDJET++ model also works at RHIC energies and its substantiation can be seen from Fig. 8 in Ref. [21].

TABLE I. Model parameters for Xe-Xe collisions at 5.44 TeV in each centrality.

| Input parameter | Value |
|---|---|
| $T_{\text{ch}}$ | 151 MeV |
| $T_{\text{th}}$ | 105 MeV |
| $\mu_{\text{th}}$ | 0 |
| $\mu_{\text{B}}$ | 0 |
| $\mu_{\text{S}}$ | 0 |

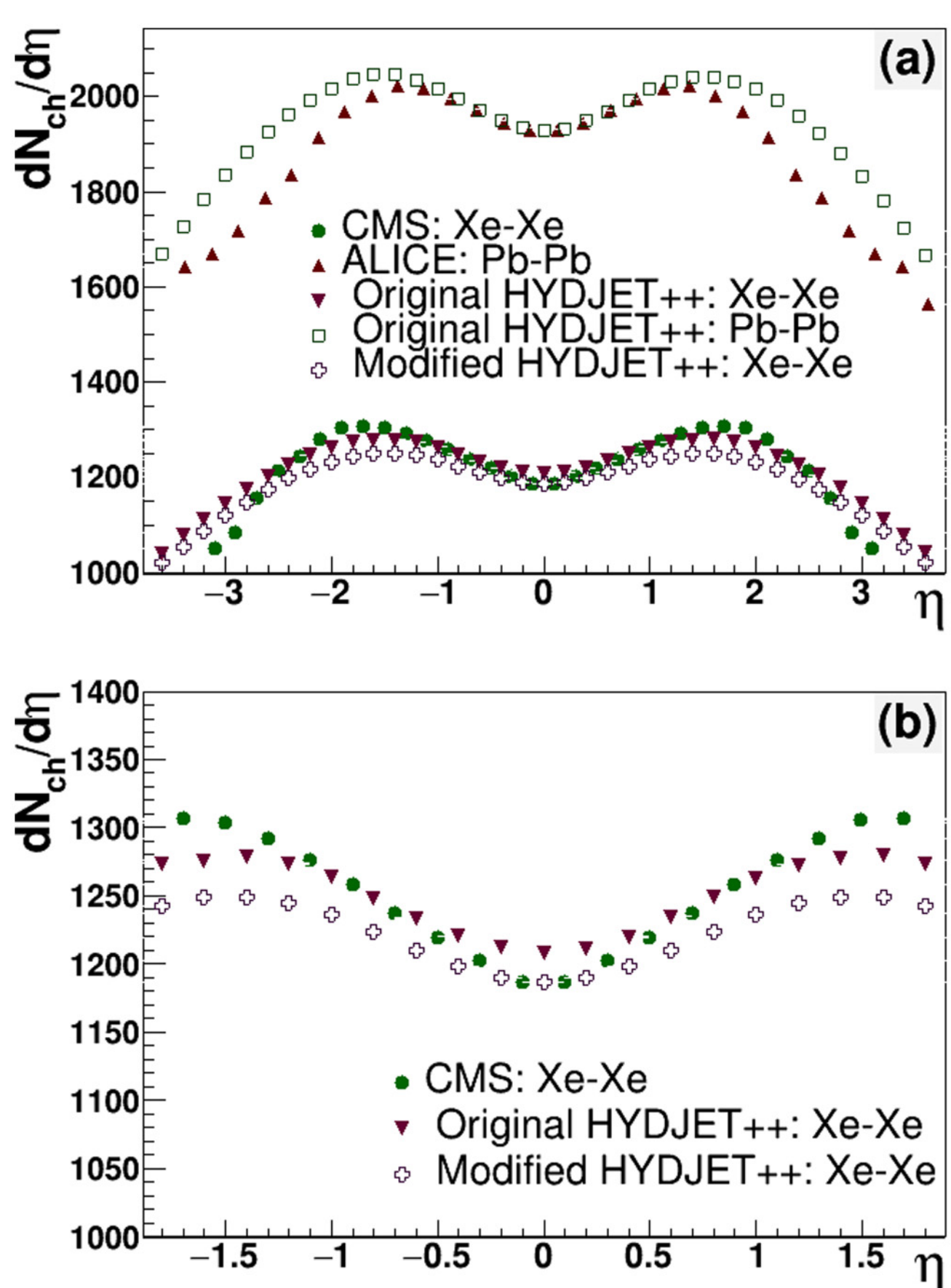

FIG. 4. (a) Pseudorapidity distribution of all charged particles in most central (0–5)% class of Xe-Xe collisions at 5.44 TeV and Pb-Pb collisions at 5.02 TeV along with the experimental data of the CMS [44] and ALICE [45] Collaborations, respectively. (b) A closer and clear view of part (a) in the midrapidity region for xenon.

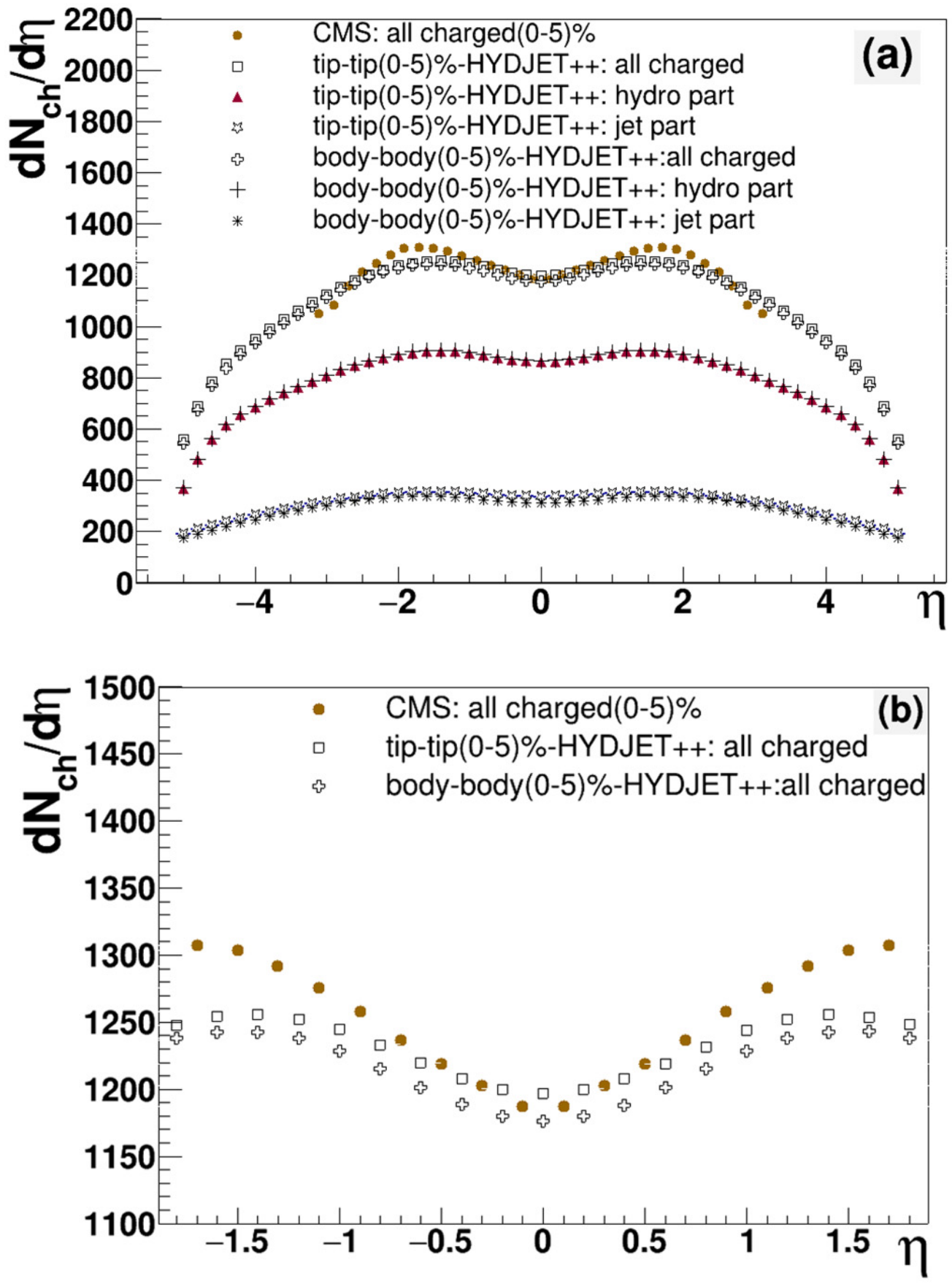

FIG. 5. (a) Pseudorapidity distribution of all charged particles in most central (0–5)% classes of Xe-Xe collisions at 5.44 TeV along with the experimental data of the CMS Collaboration [44]. (b) A closer and clear view of part (a) in the midrapidity region.

Figure 5 shows two parts. Part (a) presents the pseudorapidity distribution of charged hadrons produced in most central (0–5)% Xe-Xe collisions at 5.44 TeV. We have compared our model results with the experimental data of the CMS Collaboration [44]. Part (b) shows a clear and closer view of part (a) in the midrapidity region. We find that our model describes the shape of the distribution in both tip-tip and body-body geometrical configurations. The soft part contributes more to particle production than the hard part. The hydro part is almost three times the jet part. The soft part makes a major contribution to the central dip at $\eta \approx 0$. From the closer view in part (b), we find that the total multiplicity (hydro + jet) is higher in the tip-tip configuration compared to the body-body configuration. The hydro part shows an almost equal multiplicity in both configurations whereas the jet part has higher multiplicity in the tip-tip configuration than the body-body configuration. However, quantitatively our model fails to produce the total multiplicity at higher rapidities. This is because the HYDJET++ model incorporates Bjorken boost invariant hydrodynamics which does not work much at larger rapidities.

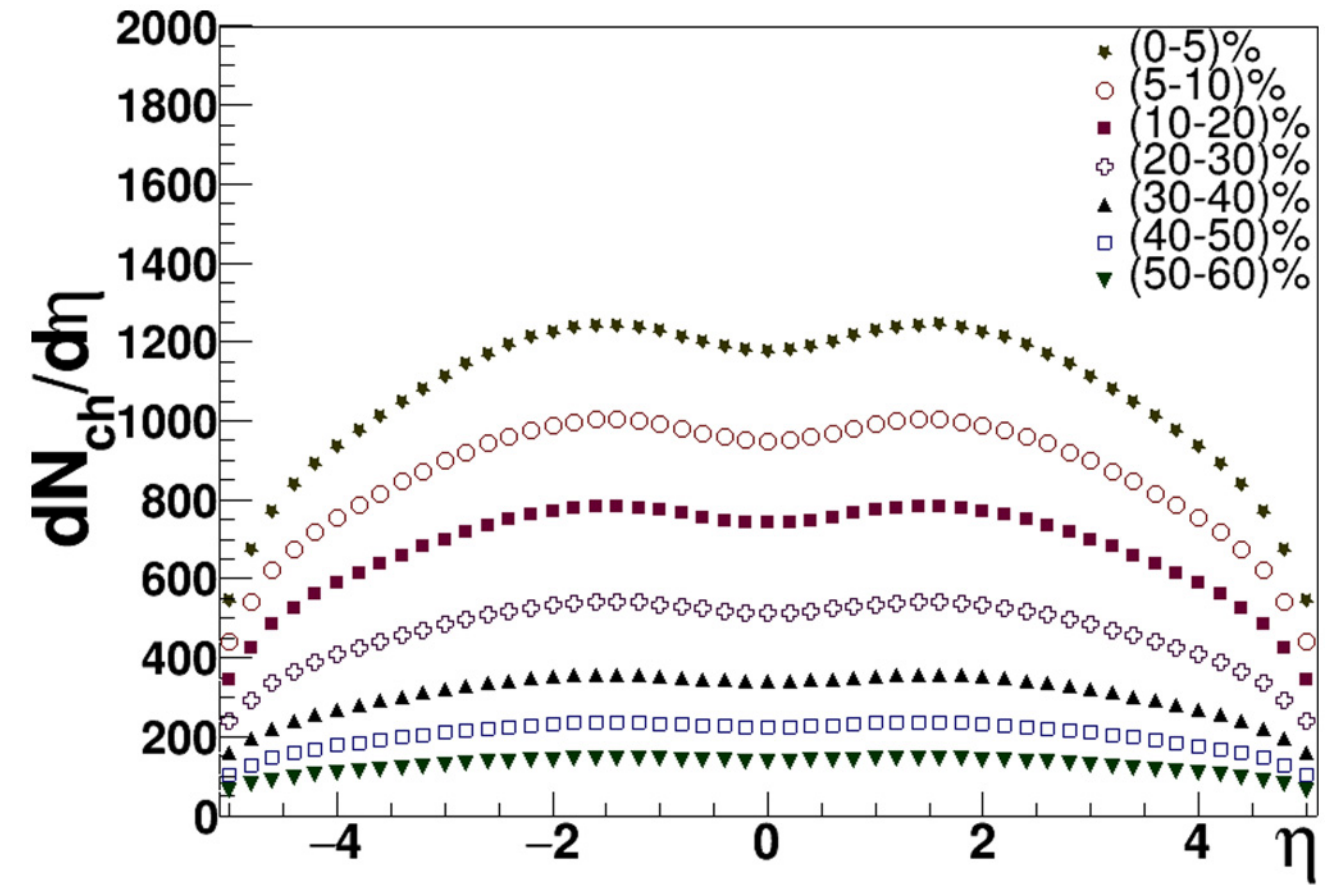

FIG. 6. Variation of $dN_{\rm ch}/d\eta$ with respect to $\eta$ of all charged particles in the body-body configuration of Xe-Xe collisions over seven classes of centrality under the HYDJET++ framework.

Figure 6 depicts the variation of $dN_{\rm ch}/d\eta$ with respect to $\eta$ in the body-body configuration. The results have been shown from most central (0–5)% class of collisions to most peripheral (50–60)% classes of collisions. Each distribution in the figure has a peak at $|\eta| \approx 1.6$ with a central dip at $\eta = 0$. The peak value of the charged hadrons is approximately 1170 for most central collisions whereas around 130 in most peripheral collisions. Hence, as we move from the most central to the most peripheral class of collisions, the number of charged hadrons produced at midrapidity decreases nine times. Similarly, Fig. 7 depicts the variation of $dN_{\rm ch}/d\eta$ with respect to $\eta$ in tip-tip collisions over all classes of centrality. Here again we find a peak at $|\eta| \approx 1.6$ with a central dip at $\eta = 0$. The peak value of the charged hadrons in tip-tip collisions is approximately 1200 for most central collisions whereas around 100 in most peripheral collisions. Thus, the total multiplicity of charged hadrons at midrapidity decreases almost 12 times as we move from the most central to the most peripheral class of collisions.

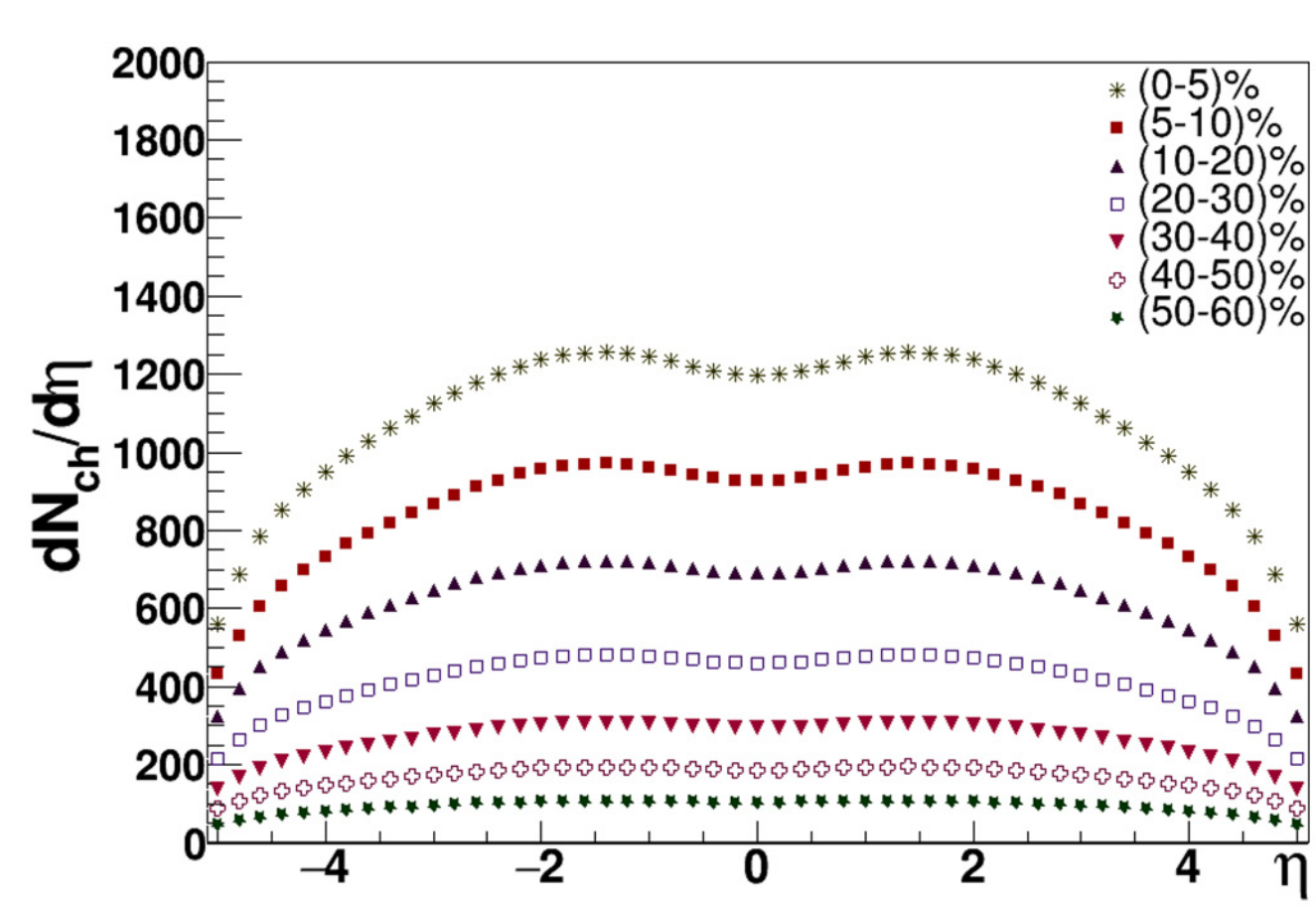

FIG. 7. Variation of $dN_{\rm ch}/d\eta$ with respect to $\eta$ of all charged particles in the tip-tip configuration of Xe-Xe collisions over seven classes of centrality under the HYDJET++ framework.

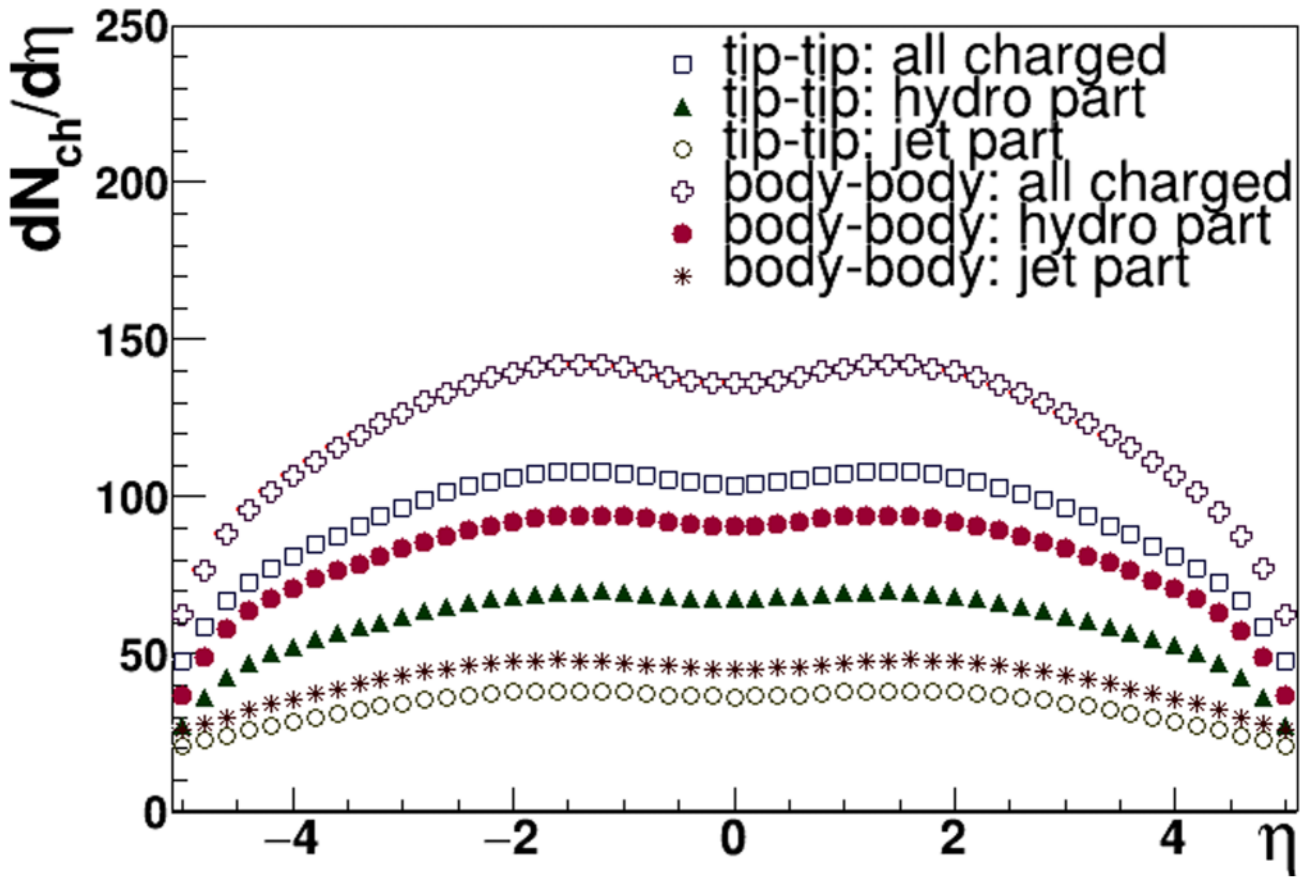

FIG. 8. Pseudorapidity distribution of all charged particles in most peripheral (50–60)% classes of Xe-Xe collisions at 5.44 TeV under the HYDJET++ framework.

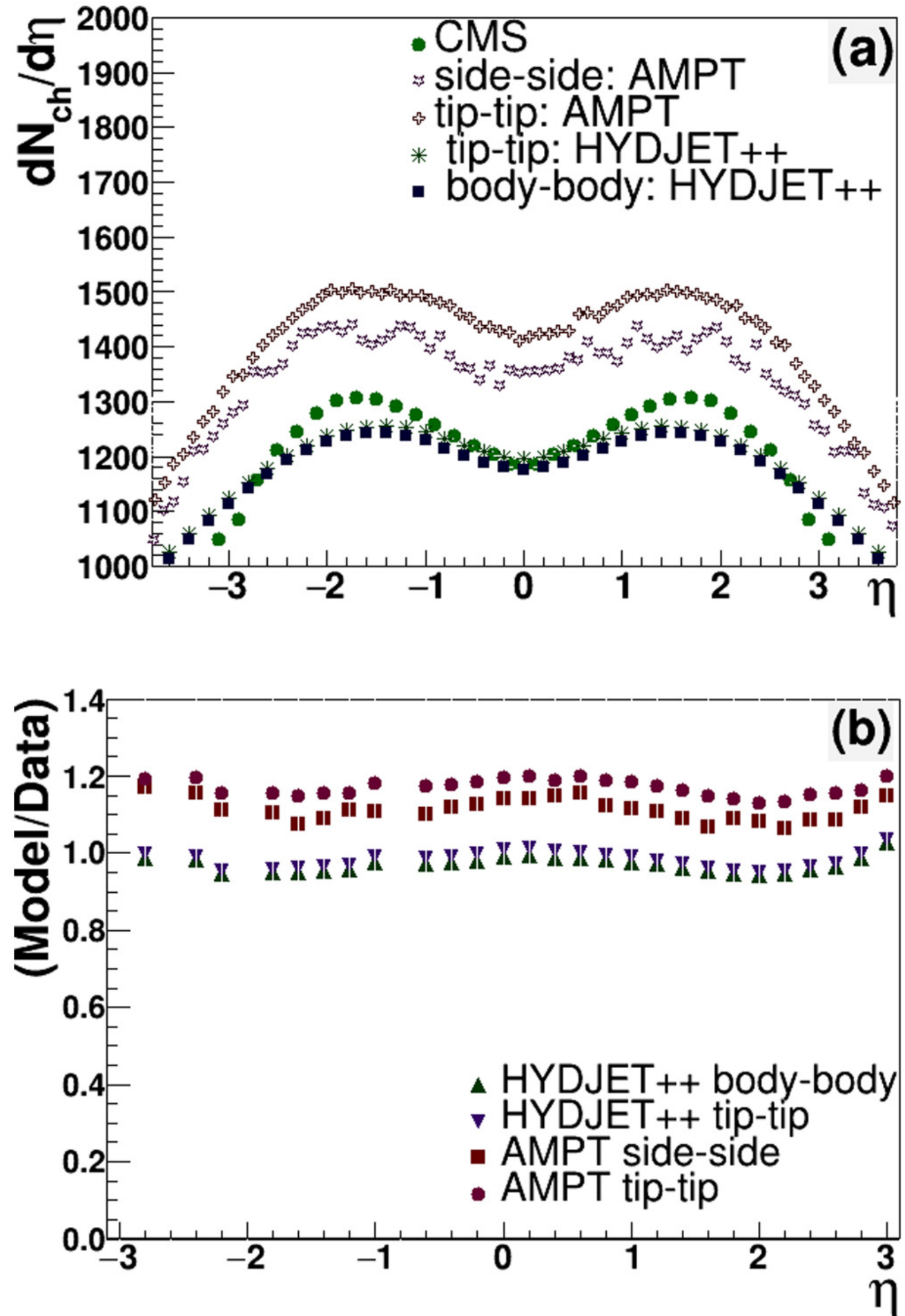

FIG. 9. (a) Comparison of $dN_{\mathrm{ch}}/d\eta$ as a function of $\eta$ of all charged particles in most central (0–5)% classes of collisions to AMPT model predictions [16] along with the experimental data of the CMS Collaboration [44]. (b) Ratio of model results to experimental data [44] for most central (0–5)% classes of collisions.

Figure 8 presents the pseudorapidity distribution of charged hadrons produced in most peripheral (50–60)% Xe-Xe collisions at 5.44-TeV center-of-mass energy. Here we have compared our body-body configuration results with tip-tip collision results. We see that the total multiplicity (soft + hard) is higher in body-body collisions. Also, the multiplicity for the jet part and the hydro part is higher in body-body collisions. The hydro part gives a larger contribution to the total multiplicity than the jet part, the hydro part being almost twice the jet part. Again, the soft part gives a greater contribution to the central dip at $|\eta| = 0$ than the hard part. Thus, from Fig. 5 as well as from Fig. 8 we infer that the particle production in smaller systems, such as xenon, is governed not only by the initial geometrical configurations, but also by the centrality at which collisions take place. The contribution of the soft part varies with centrality which brings about the difference between the initial geometrical configurations. For most central (0–5)% collisions, we cannot differentiate between the configurations in terms of soft part whereas for most peripheral (50–60)% collisions, this difference can be easily seen where the body-body soft part is almost 1.3 times the tip-tip soft part.

Figure 9 shows two parts. The top part (a) shows $dN_{\mathrm{ch}}/d\eta$ as a function of pseudorapidity for most central (0–5)% collisions. Here we have compared our HYDJET++ model results with AMPT model results [16] and the experimental data of the CMS Collaboration. The total multiplicity for side-side and tip-tip collisions from the AMPT model are higher than body-body and tip-tip collisions from the HYDJET++ model. Also, our HYDJET++ model results show good agreement with the experimental data of the CMS Collaboration [44] at midrapidities. The difference of the total multiplicity between the side-side collisions and tip-tip collisions from the AMPT model is more than the difference of the total multiplicity between the body-body collisions and tip-tip collisions from the HYDJET++ model. Furthermore, we see that AMPT is only able to produce the shape of the distribution but it completely overpredicts experimental data, whereas our HYDJET++ model results not only produce the shape of the pseudorapidity distribution, but also the experimental data lie in between the two geometrical configurations at midrapidity. The credit here may be given to PYTHIA that performs individual nucleon-nucleon collisions at the parton level whereas the Lund strings are hadronized as an ensemble [46]. The bottom part (b) of Fig. 9 shows the ratio of the models to data. Here we have compared our calculations from theoretical models to the Xe-Xe experimental data. The AMPT model shows significant deviation from the experimental data having model by data values $\gg 1$ whereas our model results look justifiable for Xe-Xe collisions having the ratios close to 1.

Figure 10 shows the variation of $dN_{\mathrm{ch}}/d\eta$ with centrality at midrapidity. Here, we have compared our minimum bias results with the experimental data of the CMS Collabora-tion [44]. Studying how well the charged-particle production depends on the centrality of collisions throws light on the effective roles played by the hard scatterings and the soft processes upon collisions at various centrality classes [47]. We see a potential dependence of $dN_{ch}/d\eta$ at midrapidity on the centrality of collisions. The total charged-particle multiplicity at midrapidity decreases as we move from most central to most peripheral classes of collisions. Our model shows a suitable match with the experimental data within the permissible error range. We have also compared our results for tip-tip and body-body collisions. In the most central class, the value of $dN_{ch}/d\eta$ is slightly higher for tip-tip collisions whereas that of body-body collisions is least. As we move towards the peripheral class of collisions, the situation reverses. The total charged-particle multiplicity is higher for body-body collisions throughout the centrality whereas it is lower for tip-tip collisions throughout the centrality. Hence, we may conclude that our model has an overall good agreement with the data at midrapidity.

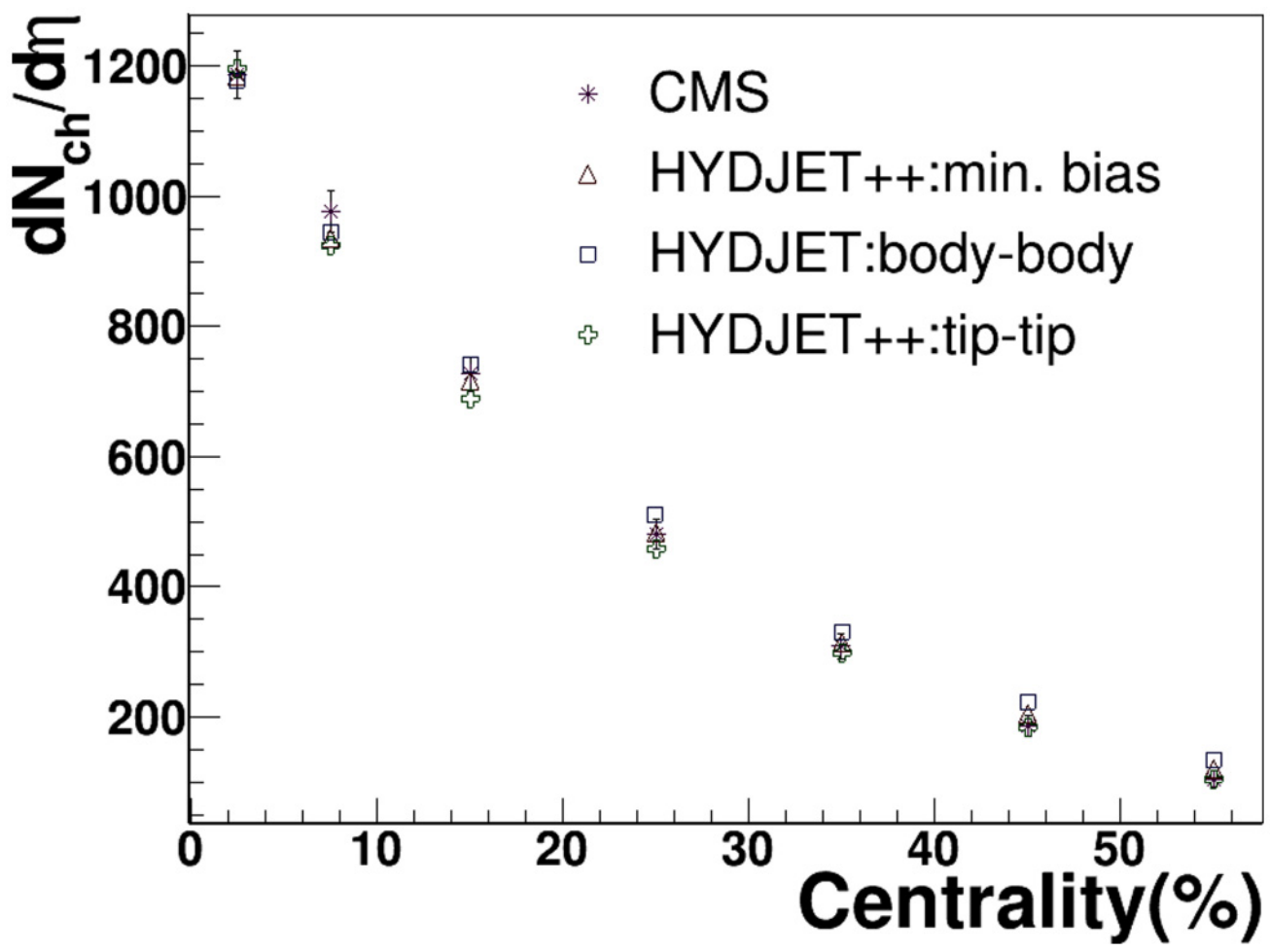

FIG. 10. Variation of $dN_{ch}/d\eta$ with centrality at midrapidity in the minimum bias configuration of Xe-Xe collisions. Also, showing the experimental data of the CMS Collaboration for comparison [44].

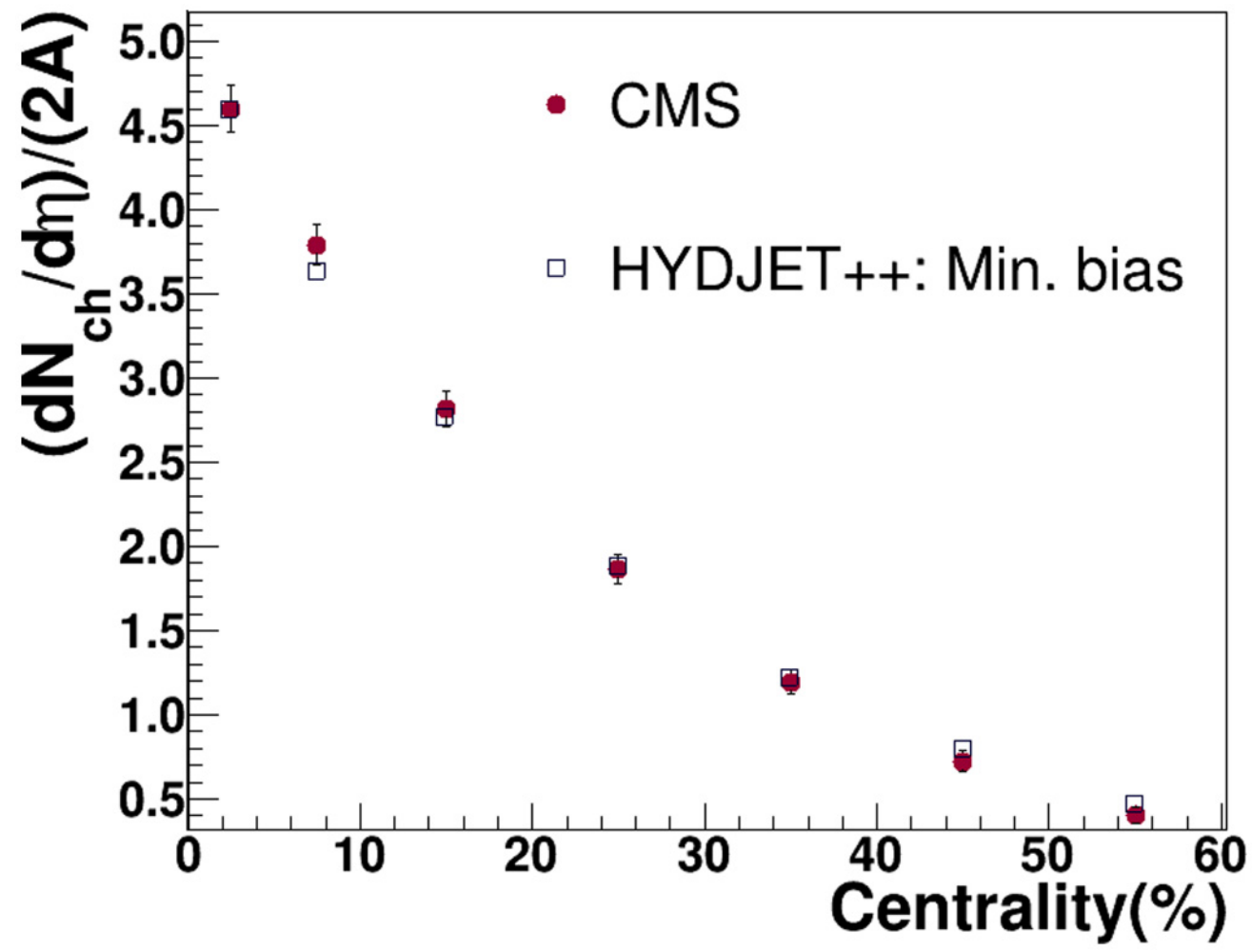

FIG. 11. Variation of charged hadron pseudorapidity density in minimum biased Xe-Xe collisions as a function of event centrality being compared with the experimental data of the CMS Collaboration [44].

Figure 11 depicts the plot for charged-hadron pseudorapidity density in minimum bias Xe-Xe collisions normalized by $2A$, where $A$ is the atomic number of the nuclei as a function of event centrality. The model results have also been compared with the experimental measurements of the CMS Collaboration [44]. A strong centrality dependence is observed here. The total charged-particle multiplicity at midrapidity shows a gradual decrease as we move from the most central to the most peripheral class of collisions. Our model matches with experimental data very well in most central collisions but underpredicts the data in midcentral collisions. Furthermore, in semiperipheral collisions, our model shows an excellent match with data whereas for most peripheral collisions it overpredicts the experimental data. The fact that the geometry of the colliding system plays a significant role in the determination of particle production [48] is not only observed at lower energies, but also observed higher energies as much as at the LHC. Thus, there exists a feasible agreement of our model results with the experimental data of the CMS Collaboration.

Furthermore, Fig. 12 represents the variation of average charged-hadron pseudorapidity density normalized by $2A$ with respect to $N_{part}$ normalized by $2A$. Results from our model have been compared with the experimental data of the CMS Collaboration [44]. The value of $dN_{ch}/d\eta$ normalized by the total number of nucleons in the two colliding nuclei strictly increases with the fraction of nucleons participating in the collisions at each centrality. In other words, the multiparticle production scales $2A$ times as a function of $N_{part}/2A$. This shows the dependence on the system size as well as on the geometry of the colliding system at which the HYDJET++

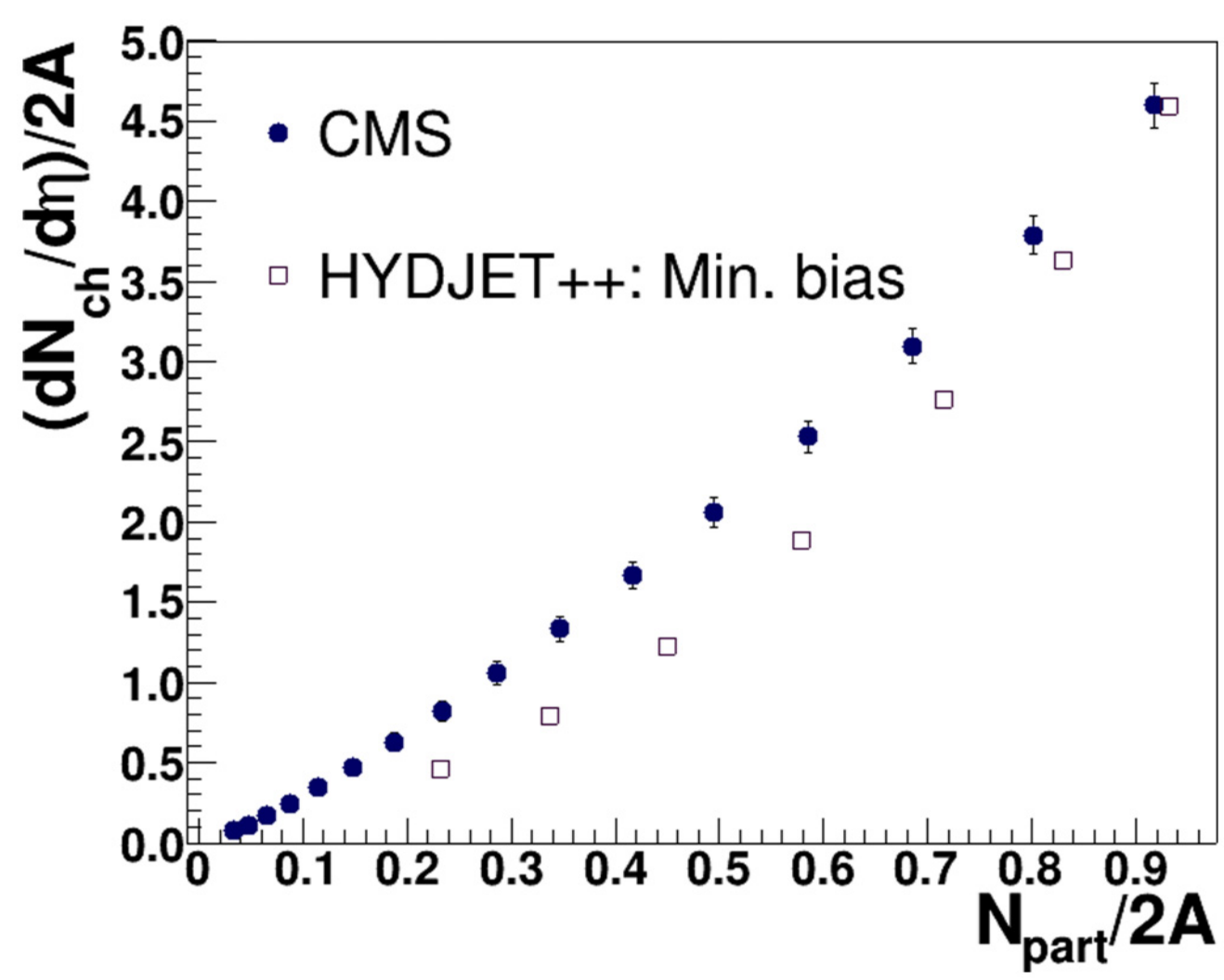

FIG. 12. Variation of the normalized average charged pseudorapidity density with respect to the number of participants ($N_{part}$) in minimum biased Xe-Xe collisions at 5.44 TeV. Also, showing the experimental data of the CMS Collaboration for comparison [44].

model is successful. Our model strongly produces the trend of experimental data, but it underpredicts the experimental measurements as we move from midcentral to peripheral collisions. The deviation from the data increases as we move towards peripheral collisions. This may be attributed to the incapability of our model to describe hard events in peripheral collisions of xenon nuclei.

### B. Transverse momentum distribution

In this section, we present the transverse momentum spectra of charged hadrons. Such spectra carry essential information about the high-density deconfined state of strongly interacting matter, quark-gluon plasma. The kinematic ranges $0.1 < p_T < 50$ GeV/$c$ and $|\eta| < 0.8$ are considered in our paper.

Figure 13 shows transverse momentum spectra of charged hadrons in two geometrical configurations of Xe-Xe collisions at 5.44 TeV. Here we have compared our model results with the experimental data of the ALICE Collaboration [49]. The figure shows three classes of centrality, namely, (0–5)%, (20–30)%, and (50–60)%. Our model results for the two geometrical configurations show a suitable match with experimental data up to $p_T \approx 4.0$ GeV/$c$ in each centrality class. So, at such $p_T$ ($p_T < 5.0$ GeV/$c$) we are unable to differentiate between the two geometrical configurations. This is also a challenging task that must be resolved. We will discuss this later. Moving further, as we move towards higher values of $p_T$ ($p_T > 5.0$ GeV/$c$), the difference between the two configurations can be visualized. In most central class of collisions (0–5)%, body-body results match with the data of the ALICE Collaboration whereas in the semiperipheral case (20–30)%, body-body results slightly overpredict the data of the ALICE Collaboration and in most peripheral (50–60)% classes of collisions, body-body results completely overpredict the data whereas tip-tip collisions completely overpredict the data of the ALICE Collaboration throughout centrality. However, in Ref. [16], AMPT model results overpredict the experimental data in central collisions for $p_T < 2.0$ GeV/$c$. It shows a suitable match in midcentral collisions and overpredicts the data for $p_T > 2.0$ GeV/$c$ in all other classes of collisions.

In most central (0–5)% classes of collisions, the slope of the $p_T$ distribution of tip-tip collisions is lower than the slope for body-body collisions. Similar behavior can be seen for (20–30)% centrality classes of collisions. This indicates that the source temperature is higher in tip-tip collisions than in body-body collisions. But as we move to most peripheral (50–60)% collisions, not much difference can be seen in the spectra of body-body and tip-tip collisions. The difference between the geometrical configurations vanishes as we move from the most central to the most peripheral class of collisions.

Figure 14 represents transverse momentum spectra of charged hadrons with respect to $p_T$ in the tip-tip configuration of Xe-Xe collisions at 5.44 TeV. Here we have obtained results for seven classes of centrality, starting from most central (0–5)% to the most peripheral (50–60)% classes of collisions. The transverse momentum spectra for each centrality class has been scaled by some weight factors in order to have a clear view of the spectrum. The slope of the $p_T$ distribution increases as we move from the most central to the most peripheral collisions. This indicates that the fireball temperature in central collisions is higher than that created in peripheral collisions.

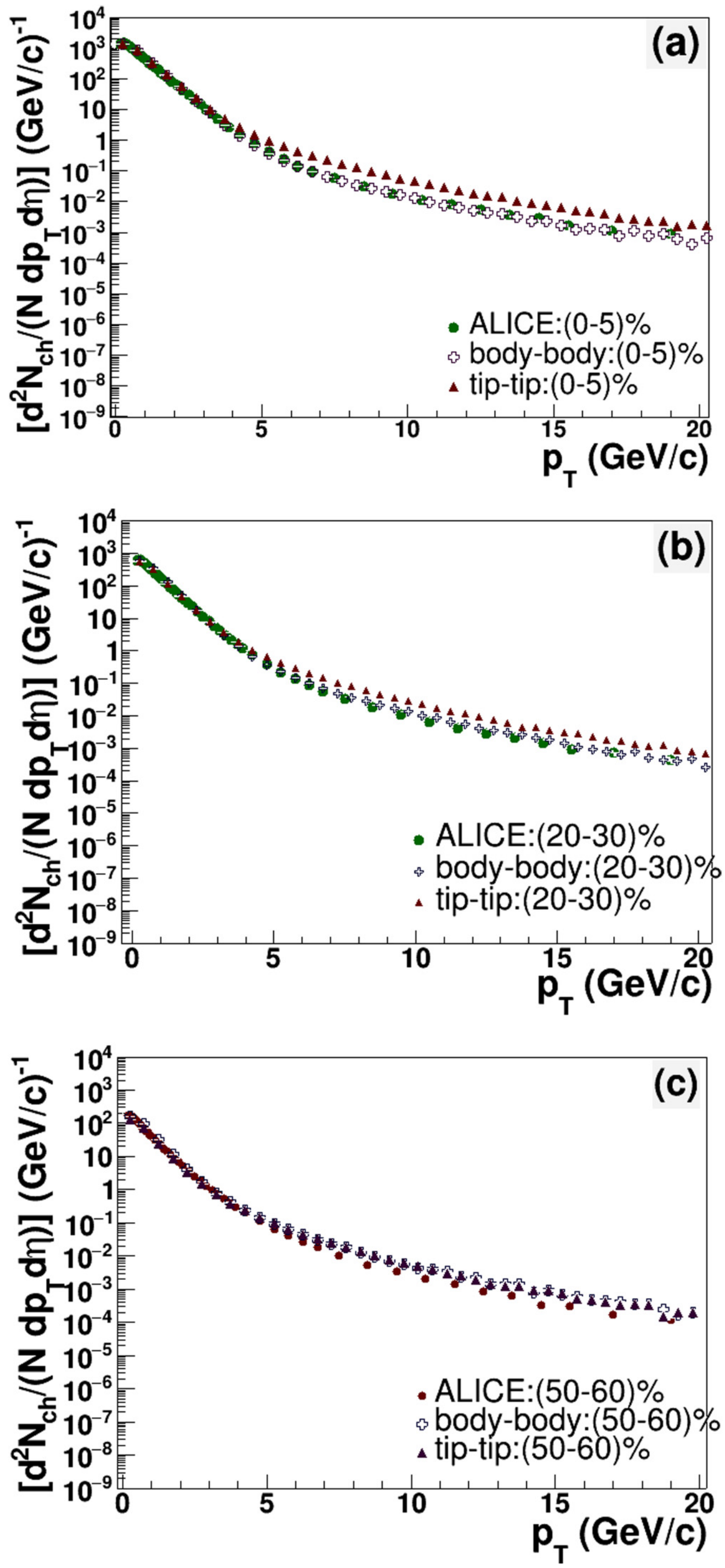

FIG. 13. Comparison of transverse momentum spectra of all charged particles in body-body and tip-tip configurations for collisions over three centralities along with the experimental data of the CMS Collaboration for comparison [49]. Here we have used a $|\eta| < 0.8$ pseudorapidity cut.

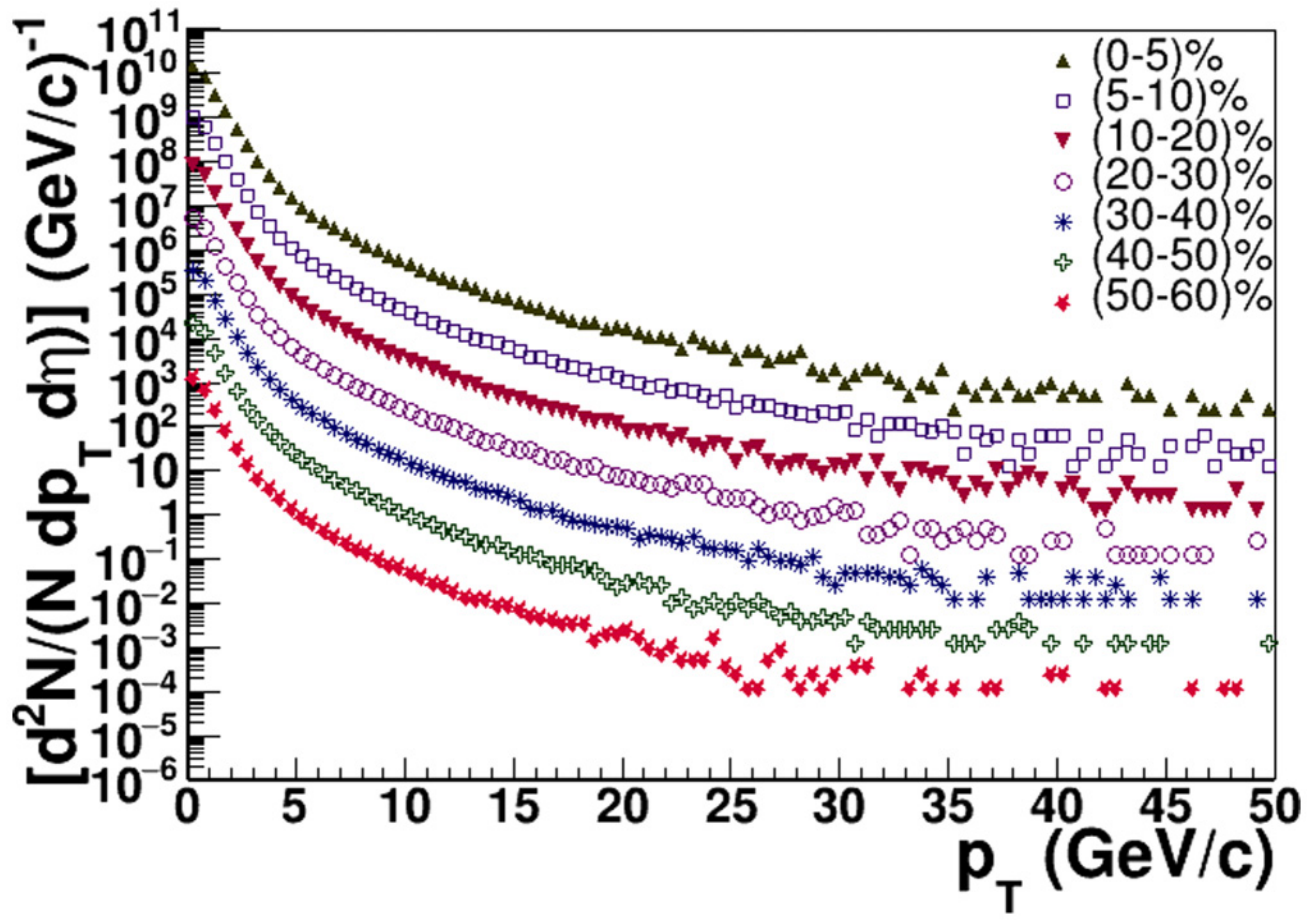

FIG. 14. Transverse momentum distribution of charged hadrons in the tip-tip configuration over seven classes of centralities.

Similarly, Fig. 15 shows the transverse momentum distribution of charged hadrons in the body-body configuration of Xe-Xe collisions at 5.44 TeV. Again here we have obtained results for seven centrality classes from the most central (0–5)% to the most peripheral (50–60)% classes of collisions. We have scaled the $p_T$ distribution of each centrality class by some weight factor for a clear view of the results. The slope of the $p_T$ distribution, which measures the inverse of the source temperature of the fireball, increases as we move from the most central to the most peripheral collisions. This again attests to the fact that the fireball temperature is higher in (most) central collisions than in peripheral collisions.

Figure 16 presents the ratio of our model results to the experimental data [49] for transverse momentum spectra in body-body and tip-tip geometrical configurations. In this way, we try to envision the difference between the two geometrical configurations. In most central (0–5)% collisions at very low values of $p_T$ ($p_T < 2.5$ GeV/$c$) the ratio is higher for tip-tip collisions than body-body collisions. However, the difference between the two is very low. As we move towards higher

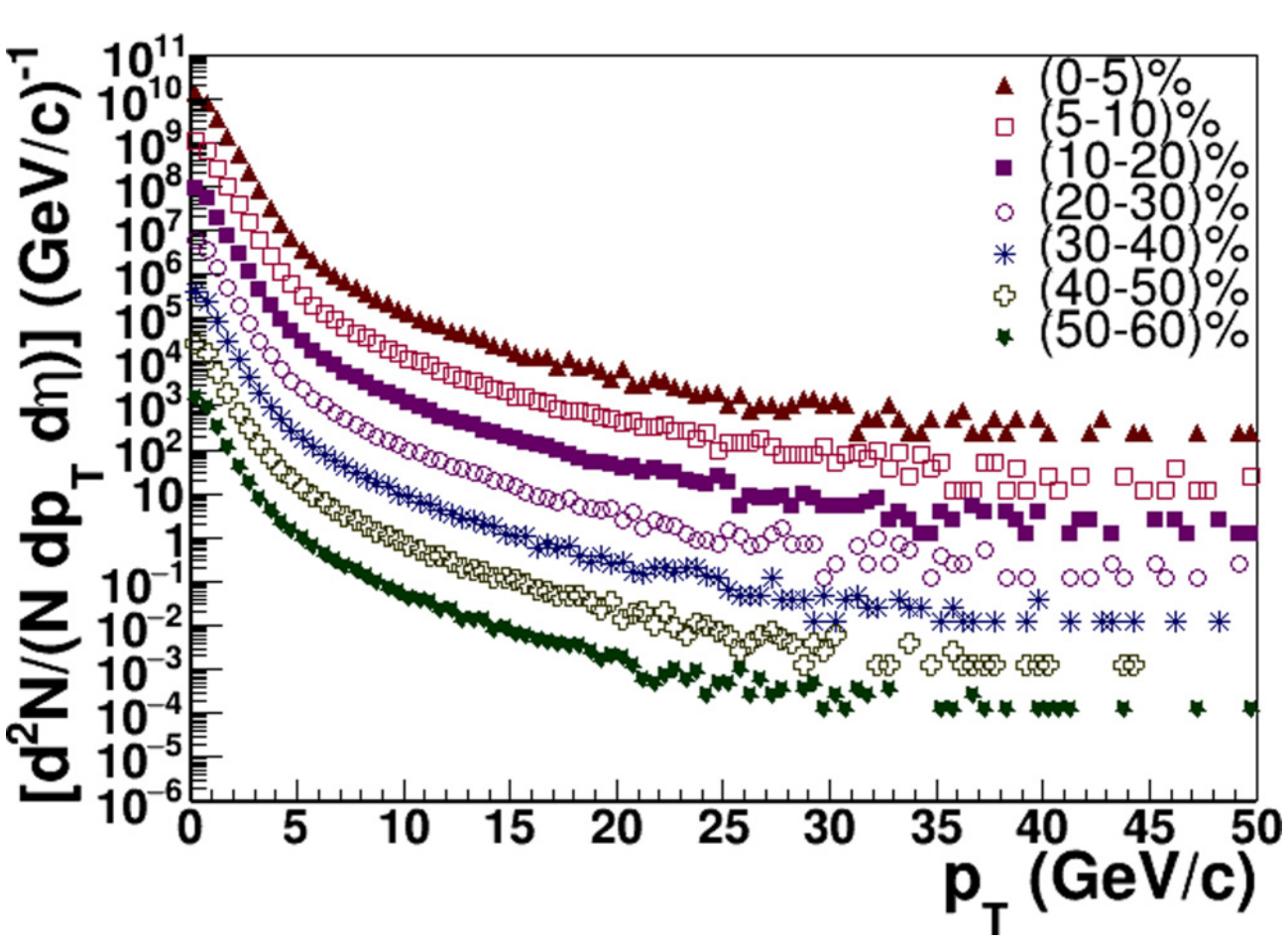

FIG. 15. Transverse momentum distribution of charged hadrons in the body-body configuration over seven classes of centralities.

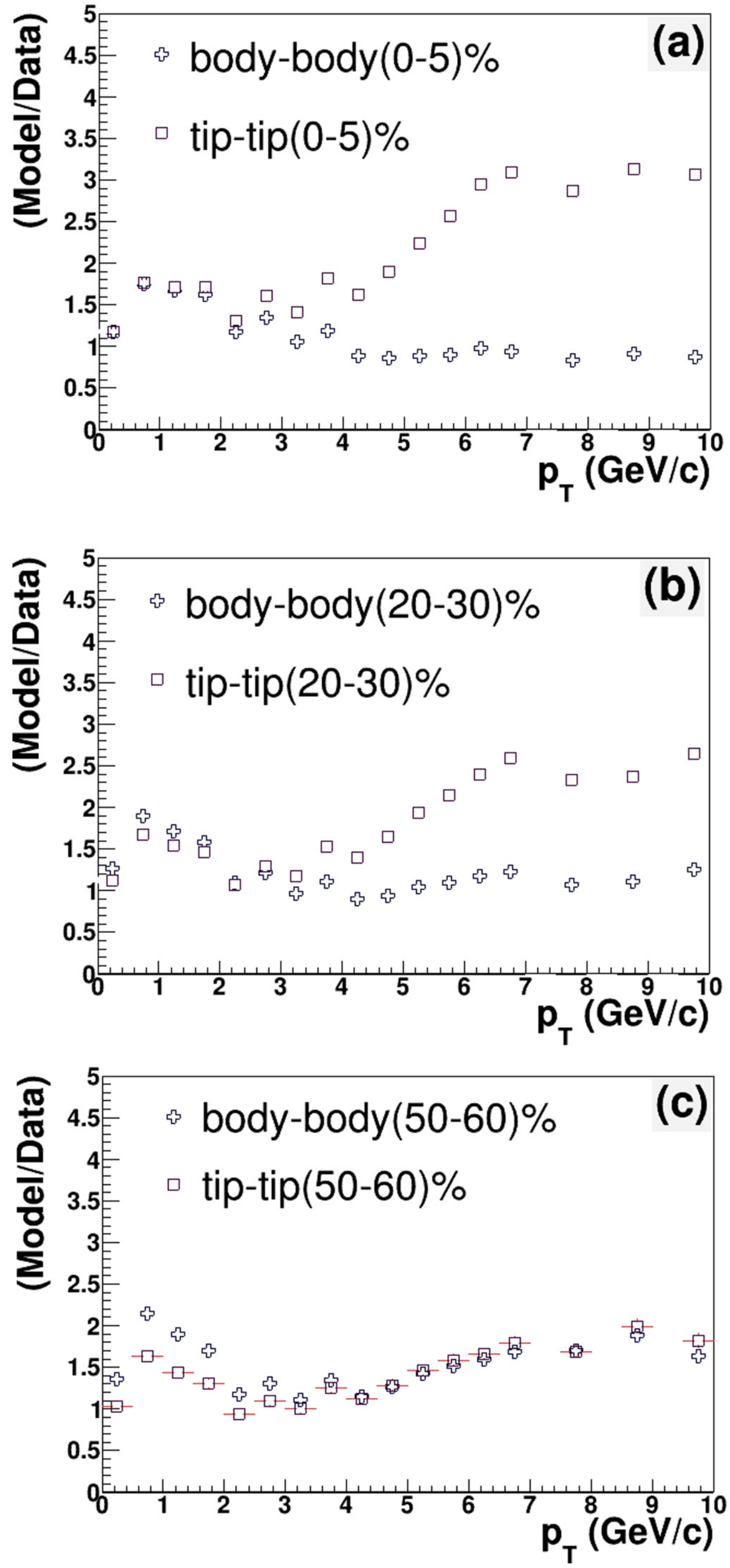

FIG. 16. Ratio of model results to experimental data [49] for body-body and tip-tip geometrical configurations.

$p_T$ values, the difference between the two configurations in terms of the ratio increases. For $p_T > 4.0$ GeV/$c$ the ratios for body-body collisions are very close to unity whereas ratios for tip-tip collisions are around 3.0. In semiperipheral (20–30)% collisions, for $p_T < 2.5$ GeV/$c$ the ratio is higher for body-body collisions, and as we move towards higher $p_T$ the situation becomes just opposite, ratios of tip-tip collisions are higher. Again as we move towards higher $p_T$ values, the difference between the two configurations in terms of ratio increases. For $p_T < 2.5$ GeV/$c$ the ratios for body-body collisions are close to 1.2 whereas ratios for tip-tip collisions are around 2.7. Furthermore, in most peripheral (50–60)% collisions, for $p_T < 2.5$ GeV/$c$ we see a clear difference between the ratios of tip-tip and body-body configurations. However, the difference fades away as we move to intermediate $p_T$ values. For $p_T < 4.0$-GeV/$c$ ratios for body-body collisions are higher whereas for $p_T > 4.0$-GeV/$c$ tip-tip collisions have higher ratios, again the difference between the two geometrical configurations being very low and ratios lie in the range of 1 to 2. Above inferences made are verified from plots (a)–(c) in Fig. 16. Clearly, for $p_T > 4.0$ GeV/$c$ the difference between the two configurations decreases as we move from the most central to the most peripheral class of collisions.

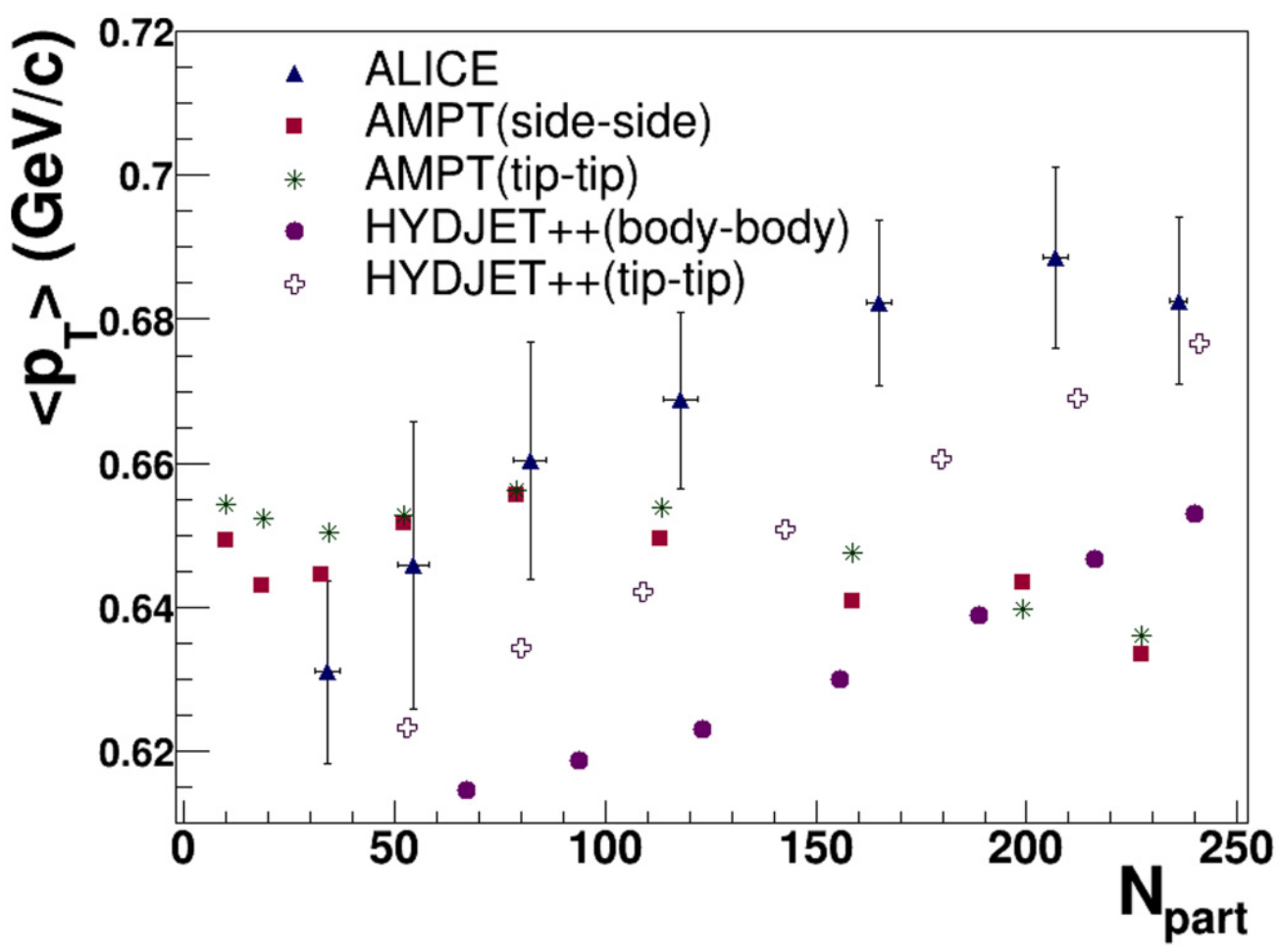

FIG. 17. Variation of average transverse momentum $\langle p_T \rangle$ with respect to $N_{\text{part}}$ at $|\eta| < 0.8$. Also showing comparison with AMPT model results [16] and the experimental data of the ALICE Collaboration [49].

In Fig. 17 we have shown the variation of average transverse momentum ($\langle p_T \rangle$) with respect to number of participating nucleons $N_{\text{part}}$ in body-body and tip-tip configurations of Xe-Xe collisions at 5.44 TeV. We present here results from the AMPT model and those from the ALICE Collaboration [49] for proper comparison. An increase in $\langle p_T \rangle$ with centrality is observed in each configuration. We find that $\langle p_T \rangle$ is smaller in our calculation than in experimental data. We see that the AMPT model [16] does not follow the trend of experimental data. Our model follows the trend of the experimental data of the ALICE Collaboration for the collisions. However, our results underpredict the data throughout centrality. As we move from the central to the peripheral collisions, $\langle p_T \rangle$ decreases. Moreover, for a particular centrality, $\langle p_T \rangle$ is higher for tip-tip collisions than for body-body collisions as expected.

### C. Elliptic flow ($v_2$)

Quark-gluon plasma, the deconfined state of color charges is believed to be created in relativistic heavy-ion collisions [50]. The pressure gradients generated in the QGP medium via most probably noncentral collisions at high relativistic energies convert the initial anisotropies to momentum anisotropies of the produced particles via multiple interactions. This phenomenon is called the anisotropic flow. This anisotropic flow is defined by the coefficients from the Fourier expansion of the azimuthal distribution of the produced particles [7,51] as

$$\frac{dN}{d\phi} \propto 1 + 2\sum_{n=1}^{\infty} v_n \cos[n(\psi - \psi_R)], \tag{4}$$

where $\psi$ = azimuthal angle of the produced particle, $n$ = harmonic value, and $\psi_R$ = reaction plane.

For $n = 2$, we have the second-order coefficient called the elliptic flow, which is sensitive to the early evolution of the system.

In the HYDJET++ model, the reaction plane is zero for each event. Here $v_2$ in terms of particle momenta is characterized by

$$v_2 = \left\langle \frac{p_x^2 - p_y^2}{p_x^2 - p_y^2} \right\rangle = \left\langle \frac{p_x^2 - p_y^2}{p_T^2} \right\rangle. \tag{5}$$

In the HYDJET++ model, soft particle emission which contributes most to the elliptic flow, comes from the freeze-out hypersurface. Figure 18 shows the variation of elliptic flow ($v_2$) with respect to $p_T$ in two geometrical configurations of Xe-Xe collisions at 5.44 TeV. Here we have also compared our model results with the experimental data of the ALICE Collaboration [52]. From the results, we see that the two geometrical configurations show the same behavior qualitatively. $v_2$ is larger for the body-body configuration as compared to the tip-tip configuration in each centrality. Our results show a suitable match with the experimental data of the ALICE Collaboration for $p_T < 1$ GeV/$c$. The magnitude of $v_2$ is least in most central (0–5)% collisions and increases as we move towards peripheral collisions. However, in most peripheral (50–60)% collisions, the magnitude of $v_2$ decreases. This is attributed to the fact that although spatial anisotropy may be maximum in peripheral collisions, the medium density is small enough to bring about any flow effect. Then, in each class of centrality, the difference between the elliptic flow of the body-body and tip-tip configurations increase as we move towards larger $p_T$. Furthermore, the maxima in $v_2$ is found to be at $p_T \approx 3.25$ GeV/$c$ in most central collisions whereas the maxima in most peripheral collisions is found at $p_T \approx 1.75;$GeV/$c$. Thus, as we move from the most central to the most peripheral collisions, the value of $p_T$ at which maxima are obtained decreases.

In Fig. 19, we discuss elliptic flow as a function of $p_T$ over seven classes of centrality in the body-body configuration of Xe-Xe collisions at 5.44 TeV. The magnitude of $v_2$ increases as we move from the most central to the most peripheral collisions. In most central collisions, $v_2$ increases and becomes maximum at $p_T \approx 3$ GeV/$c$ and then decreases as $p_T$ further increases. The value of $p_T$ at which maxima is obtained in each centrality class decreases as we move from the most central to the most peripheral collisions. However, we do not see much difference between $v_2$ of (40–50)% and (50–60)% centralities. This might be because of the formation of a smaller system in peripheral collisions. Also, xenon does not have much deformation to bring about a very large flow at peripheral collisions. At higher values of $p_T$, we cannot make much difference between $v_2$ of the various centrality collisions. Similarly, Fig. 20 shows elliptic flow distribution with $p_T$ in the tip-tip configuration of Xe-Xe collisions. The qualitative behavior is similar to that of body-body collisions as shown in Fig. 19. However, quantitatively $v_2$ is lesser in tip-tip collisions than that in body-body collisions.

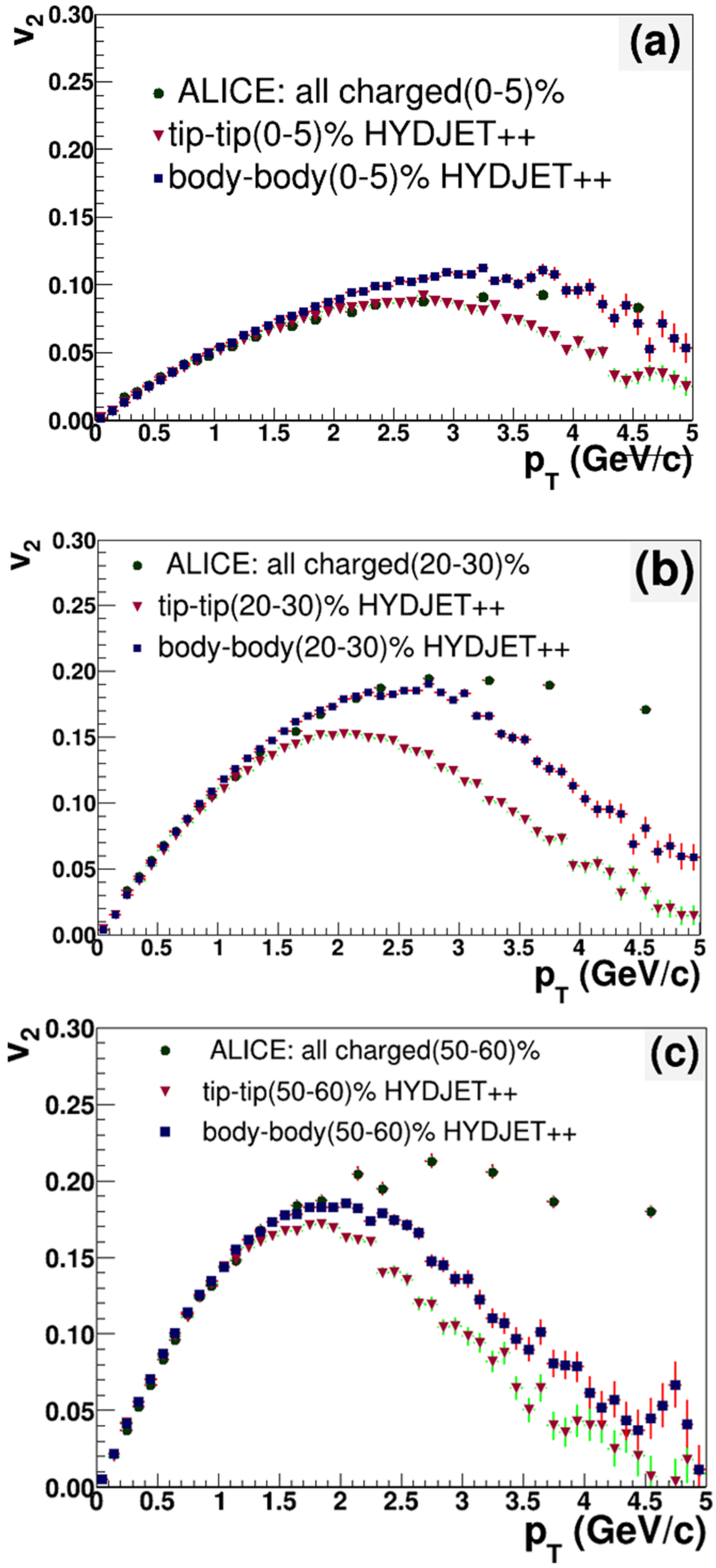

FIG. 18. Elliptic flow $v_2$ of charged hadrons with respect to $p_T$ for three classes of centrality in various geometrical configurations along with the experimental data of the ALICE Collaboration [52].

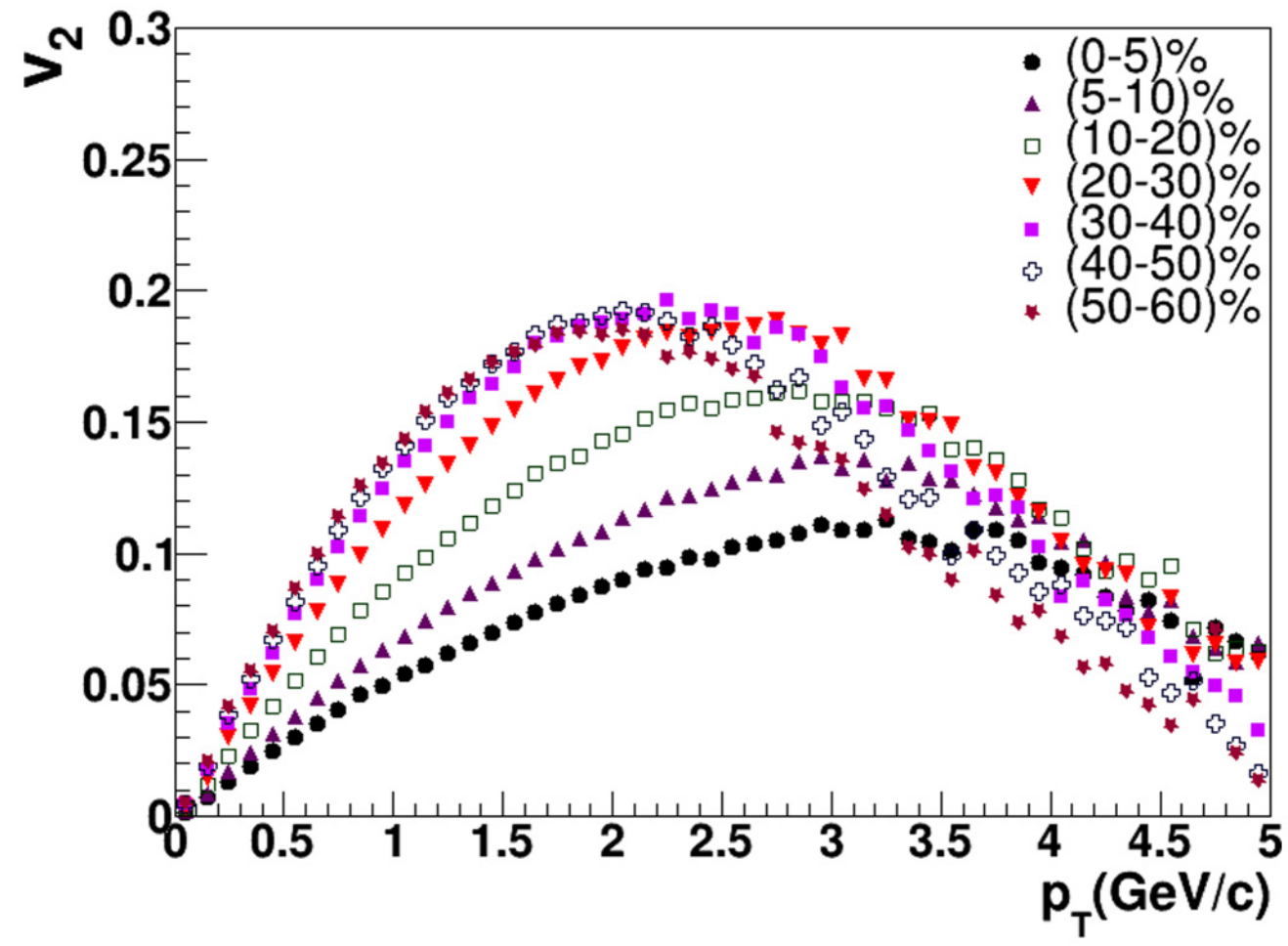

FIG. 19. Elliptic flow $v_2$ distribution of all charged hadrons in the body-body configuration over seven centrality classes.

Figure 21 presents the variation of $v_2$ with respect to $p_T$ for body-body and tip-tip collisions of xenon nuclei. The model results for these configurations have been compared to AMPT model results in side-side and tip-tip configurations [16]. Also, we have compared these results with the experimental data of the ALICE Collaboration [52]. We have shown results for four centralities (0–5)%, (10–20)%, (20–30)%, and (30–40)%. In central collisions, AMPT model completely overpredicts the experimental data of the ALICE Collaboration where the side-side configuration values being larger than that of the tip-tip configuration. However, the HYDJET++ model not only produces results qualitatively, but also shows good agreement with the data of the ALICE Collaboration quantitatively. Elliptic flow is larger in the body-body configuration than in the tip-tip configuration. The experimental data lie approximately between body-body and tip-tip collision results, evident from the (b) and (c) plots in the figure. As we move to midcentral to peripheral collisions, our body-body collision results show a close match to the data. A notable point from the results is that as we move from central to peripheral collisions, the difference between the two configurations increases. However, this is not very clear in results from the AMPT model.

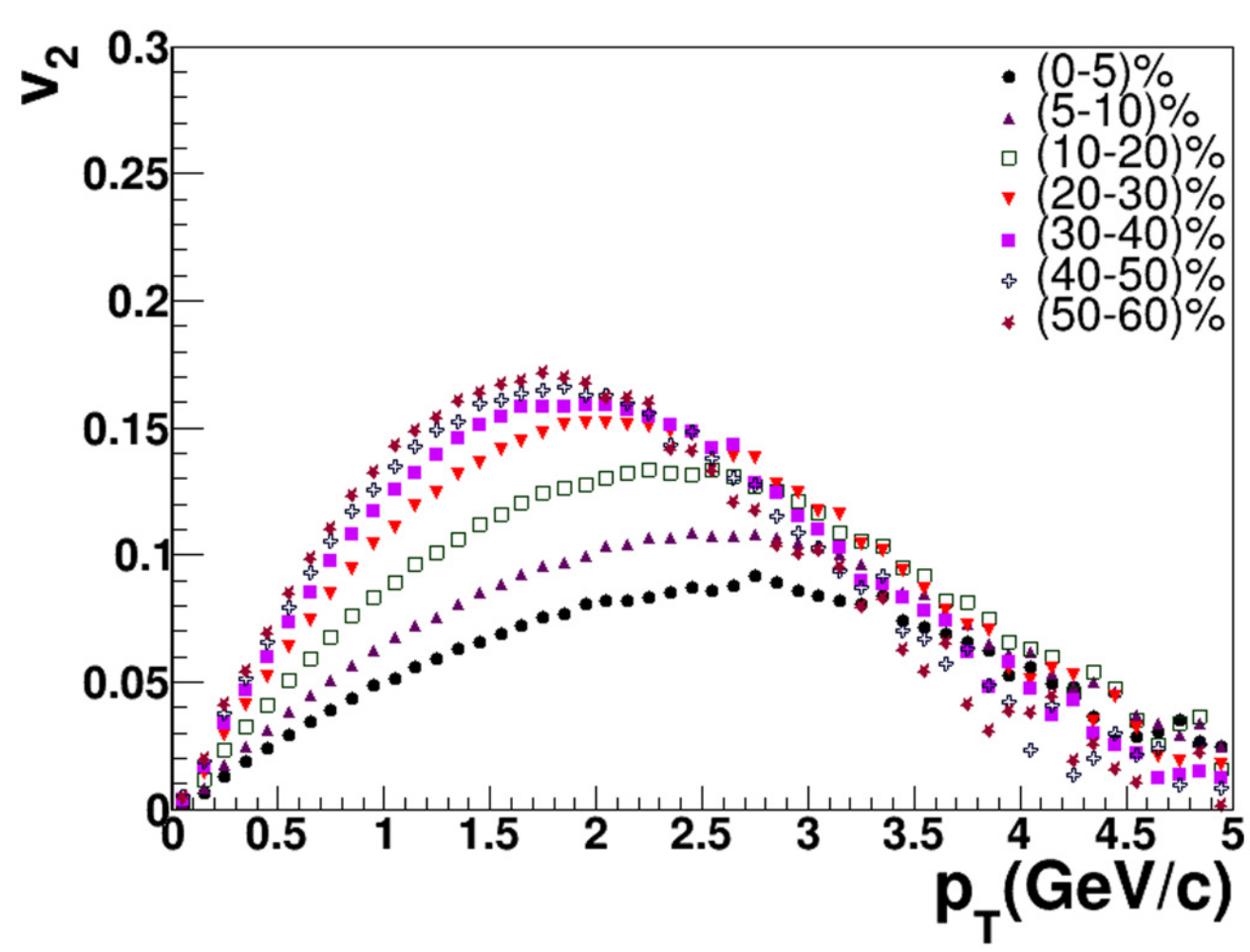

FIG. 20. Elliptic flow $v_2$ distribution of all charged hadrons in the tip-tip configuration over seven centrality classes.

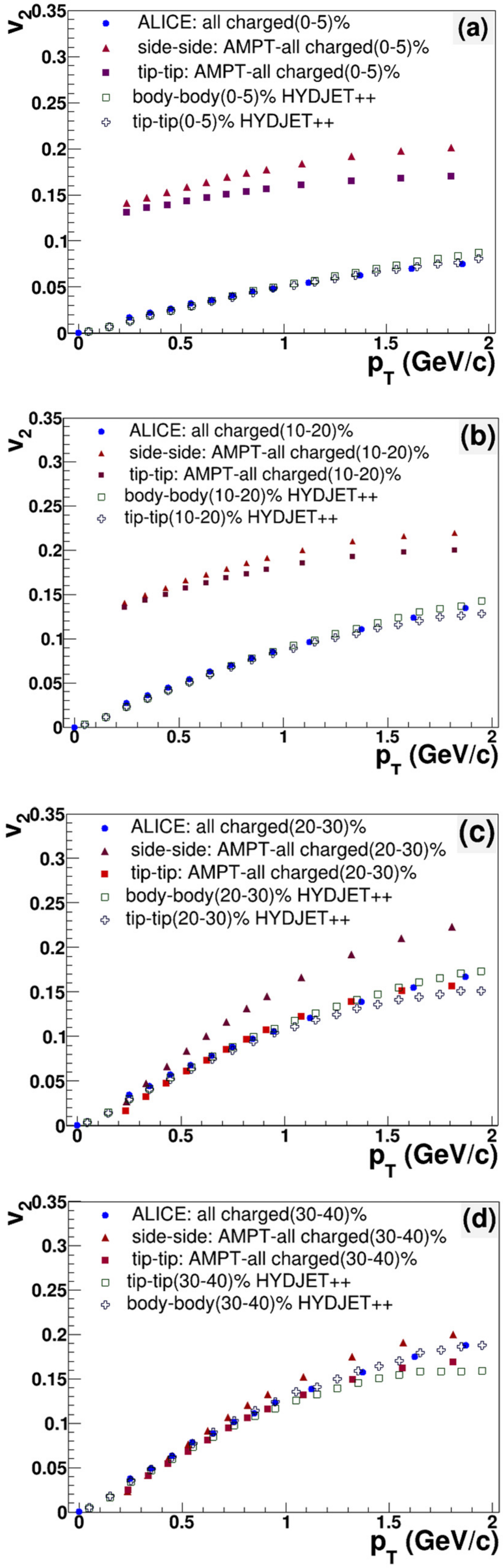

FIG. 21. Variation of elliptic flow $v_2$ with respect to $p_T$ in various centrality classes. Also showing AMPT model results [16] in various geometrical configurations along with the experimental data of the ALICE Collaboration [52] for comparison.

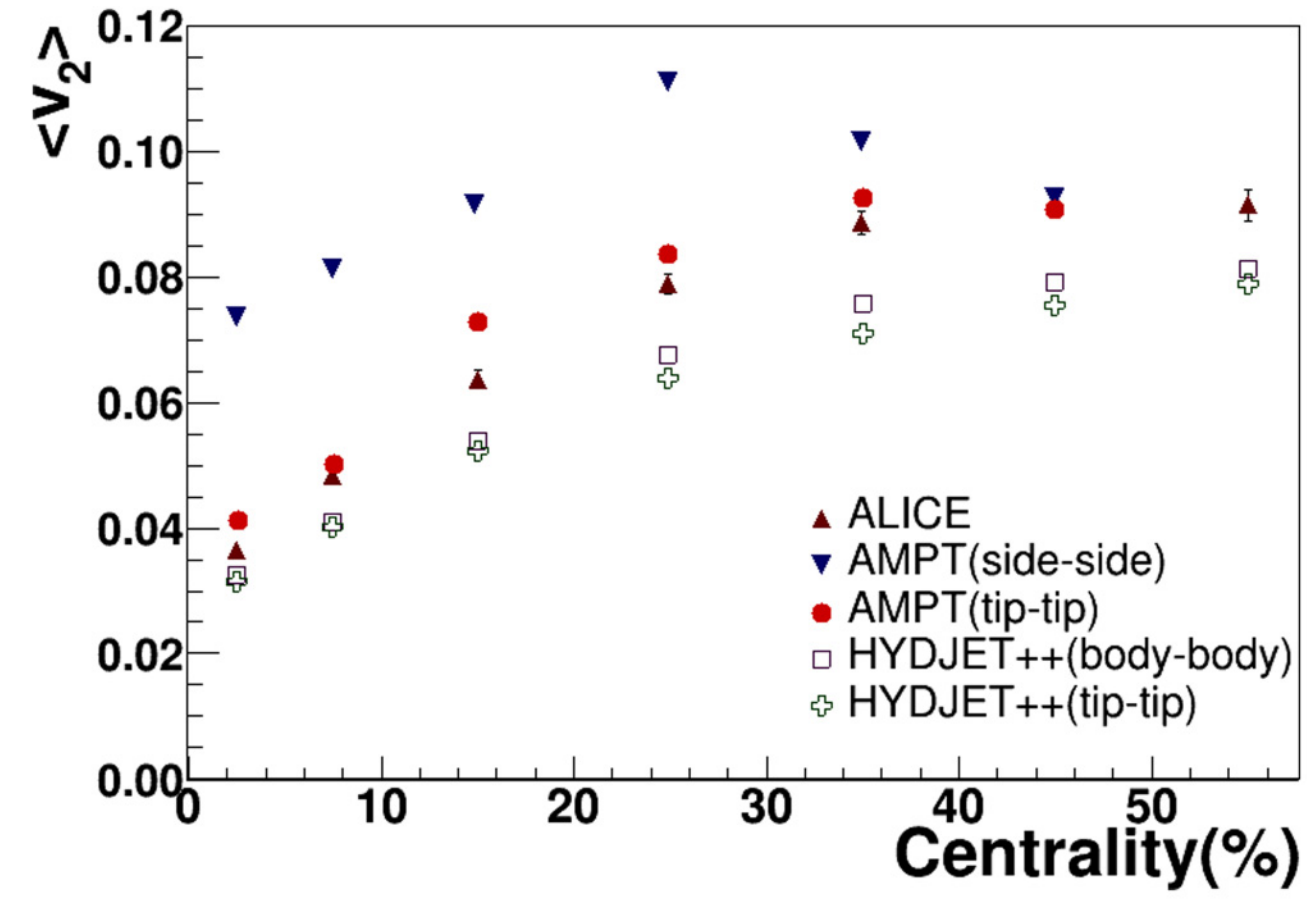

FIG. 22. Variation of average elliptic flow $\langle v_2 \rangle$ with centrality. Also comparing AMPT model results [16] in various geometrical configurations along with the experimental data of the ALICE Collaboration [52].

Lastly, we discuss the variation of average elliptic flow, $\langle v_2 \rangle$ with respect to centrality for different geometric configurations in Xe-Xe collisions at 5.44 TeV as shown in Fig. 22. A clear and mild dependence of $\langle v_2 \rangle$ with centrality is observed. Again, we have compared our model results to the results from the AMPT model [16]. The AMPT model shows a strong dependence of $\langle v_2 \rangle$ with centrality. Quantitatively, our model results underpredict the experimental measurements of the ALICE Collaboration [52] throughout centrality whereas the AMPT model overpredicts the data except for the most peripheral collisions where the results match the experimental data of the ALICE Collaboration. Qualitatively, our model results are commendable for both the geometrical configurations whereas the AMPT model results differ completely for side-side and tip-tip configurations. Furthermore, $\langle v_2 \rangle$ increases as we move from central to semiperipheral collisions and shows a decrease as we move towards most-peripheral collisions.

## IV. SUMMARY AND OUTLOOK

In a brief view of our paper, we have tried to make a scrupulous study of xenon-xenon collisions at $\sqrt{s_{NN}} = 5.44$-TeV LHC energies. Performing this in the framework of the HYDJET++ model, provides us the possibility to study the collisions in various geometrical configurations. These geometrical configurations are cognizant of the initial conditions. Here, we have used tip-tip and body-body configurations for our analysis. Also, we have used the results from the AMPT model in the string-melting version for comparison along with the experimental data of the ALICE/CMS Collaborations. We find that our model results show a suitable match with the experimental data whereas results from the AMPT model completely overpredict it.

The pseudorapidity density for minimum bias Xe-Xe collisions in our HYDJET++ model shows a suitable match with the experimental measurements of the CMS Collaboration at midrapidities. Strong centrality dependence of $dN_{\mathrm{ch}}/d\eta$ at midrapidity is observed which is consistent with the data. Another important observation made is that even in such smaller collision systems, we can measure the difference between the various geometrical configurations, which is one of the many challenging tasks for the experiments, such as the ALICE and CMS Collaborations at the LHC. The total charge-particle multiplicity is higher for the tip-tip configuration in most central collisions, but as we move towards peripheral collisions, body-body configuration gives higher multiplicity. We observe a clear dependence of the particle multiplicity on the system size and on the geometry of the colliding system.

Then, we have shown the results for transverse momentum spectra at $|\eta| < 0.8$. Our results match experimental data up to $p_T \approx 4.0$ GeV/$c$ and overpredict the data for higher values of $p_T$. For $p_T > 5.0$ GeV/$c$, the transverse momentum spectra is higher for tip-tip collisions than body-body collisions. However, the difference between the two geometrical configurations is larger in central collisions and vanishes in the most peripheral classes of collisions. At $p_T < 5.0$ GeV/$c$, central collisions have negligible differences whereas significant difference prevails in peripheral collisions. Thus, here $p_T$ comes out to be a good tool to disentangle the tip-tip collisions from body-body collisions. $\langle p_T \rangle$ shows absolute dependence on centrality. The HYDJET++ model underpredicts the experimental measurements quantitatively, but it is successful in producing the results qualitatively. AMPT model predictions are not reliable. It overpredicts the data in the peripheral class of collisions whereas underpredicts in central collisions. Hence, it fails to produce data quantitatively.

Lastly, the elliptic flow results for the geometrical configurations match experimental data up to $p_T < 1.0$ GeV/$c$. The elliptic flow measurements reveal that $v_2$ for body-body collisions is larger than $v_2$ for tip-tip collisions throughout centrality. $v_2$ increases as we move from the most central to the semiperipheral collisions and then decreases again in most peripheral collisions. The difference between the geometrical configurations increases as we move towards peripheral collisions. The AMPT model highly overpredicts the data in central collisions. The tip-tip collisions from the AMPT model match experimental data in peripheral collisions whereas side-side collisions still overpredict it. Average elliptic flow, $\langle v_2 \rangle$ also shows a smooth dependence on centrality. $\langle v_2 \rangle$ increases as we move from the most central to the semiperipheral collisions and starts to decrease as we move towards most peripheral collisions. The HYDJET++ model is successful in producing the flow behavior qualitatively but fails to produce flow quantitatively where the produced flow underpredicts the data throughout centrality. However, we are successful in disentangling the tip-tip and body-body collisions in the most peripheral class of collisions also. In comparison, the AMPT model shows strong centrality dependence. It overpredicts the data throughout centrality except for the most peripheral collisions where we can no more disentangle the geometrical configurations.

Thus, our paper will enlighten the particle production mechanism in intermediate systems, such as xenon. The effect of system size and geometry of the colliding system on such collision systems can be clearly visualized from our paper. The study of the evolution of the fireball created in various geometrical configurations of Xe-Xe collision systems reveal picture of hard and soft collisions in body-body and tip-tip configurations. The flow measurements will also illuminate our study of the chiral magnetic effect observable for our future work.

## ACKNOWLEDGMENTS

We sincerely acknowledge support from the Institutions of Eminence (IoE) BHU Grant. S.P. acknowledges the financial support obtained from UGC under a research fellowship scheme during the work. S.K.T. acknowledges the financial support from CRS Projects No. 1-5748317158 and No. 1-5730324531 funded by the All India Council for Technical Education (AICTE), Government of India.